\pdfoutput=1
\documentclass[utf8]{scrartcl} 
\usepackage{graphicx}
\usepackage{url,microtype,subcaption}
\usepackage[hidelinks]{hyperref}
\usepackage[TS1, T1]{fontenc}
\usepackage[numbers, sort&compress]{natbib}

\usepackage{glossaries-extra}
\setabbreviationstyle[acronym]{long-short}
\GlsXtrEnableEntryCounting{acronym}{3}

\usepackage{romannum}
\usepackage{siunitx}
\usepackage[inline]{enumitem}
\usepackage{authblk}

\title{Recent Advances and Prospects in the Research of Nascent Adhesions} 
\author[1]{Henning Stumpf}
\author[2]{Andreja Ambriovi\'c-Ristov}
\author[3]{Aleksandra Radenovic}
\author[1,4]{Ana-Sun\v{c}ana Smith}
\affil[1]{PULS Group, Institute for Theoretical Physics, Interdisciplinary Center for Nanostructured Films, Friedrich-Alexander-Universit\"at Erlangen-N\"urnberg, Erlangen, Germany}
\affil[2]{Laboratory for Cell Biology and Signalling, Division of Molecular Biology, Ru\dj er Bo\v{s}kovi\'{c} Institute, Zagreb, Croatia}
\affil[3]{Laboratory of Nanoscale Biology, \'Ecole Polytechnique F\'ed\'erale de Lausanne, Lausanne, Switzerland}
\affil[4]{Group for Computational Life Sciences, Department of Physical Chemistry, Ru\dj er Bo\v{s}kovi\'{c} Institute, Zagreb, Croatia}

\newacronym{md}{MD}{molecular dynamics}
\newacronym{bc}{BC}{bent-closed}
\newacronym{ec}{EC}{extended-closed}
\newacronym{eo}{EO}{extended-open}
\newacronym{fak}{FAK}{focal adhesion kinase}
\newacronym{ecm}{ECM}{extracellular matrix}
\newacronym{thd}{THD}{talin head domain}
\newacronym{cho}{CHO}{chinese hamster ovary}
\newacronym[\glslongpluralkey={nascent adhesions}, \glsshortpluralkey={NAs}]{na}{NA}{nascent adhesion}
\newacronym[\glslongpluralkey={focal adhesions}, \glsshortpluralkey={FAs}]{fa}{FA}{focal adhesion}

\newacronym[\glslongpluralkey={transmembrane domains}, \glsshortpluralkey={TMDs}]{tmd}{TMD}{transmembrane domain}
\newacronym[\glslongpluralkey={giant unilamellar vesicles}, \glsshortpluralkey={GUVs}]{guv}{GUV}{giant unilamellar vesicle}
\newacronym[\glslongpluralkey={droplet-stabilized giant unilamellar vesicles}, \glsshortpluralkey={dsGUVs}]{dsguv}{dsGUV}{droplet-stabilized giant unilamellar vesicle}
\newacronym[\glslongpluralkey={hemidesmosomes}, \glsshortpluralkey={HDs}]{hd}{HD}{hemidesmosome}
\newacronym{tgf}{TGF}{transforming growth factor}
\newacronym{m2}{M\Romannum{2}}{myosin \Romannum{2}}

\newacronym{palm}{PALM}{photoactivated localization microscopy}
\newacronym{sptpalm}{sptPALM}{single particle tracking PALM}
\newacronym{storm}{STORM}{stochastic optical reconstruction microscopy}
\newacronym{sofi}{SOFI}{super-resolution optical fluctuation imaging}
\newacronym{sr}{SR}{super-resolution microscopy}
\newacronym{bsofi}{bSOFI}{balanced SOFI}
\newacronym{tirf}{TIRF}{total internal reflection}
\newacronym{smlm}{SMLM}{single molecule localization microscopy}
\newacronym{paint}{PAINT}{points accumulation for imaging in nanoscale topography}
\newacronym{sim}{SIM}{structured illumination microscopy}
\newacronym{tie}{TIE}{transport of intensity equation}
\newacronym{ricm}{RICM}{reflection interference contrast microscopy}
\newacronym{dods}{DODS}{dynamic optical displacement spectroscopy}
\newacronym{fcs}{FCS}{fluorescence correlation spectroscopy}
\newacronym{afm}{AFM}{atomic force microscopy}
\newacronym{qpi}{QPI}{quantitative phase imaging}
\newacronym{sted}{STED}{stimulated emission depletion}
\newacronym{resolft}{RESOLFT}{reversible saturable optical fluorescence transitions}
\newacronym{emgm}{EMGM}{expectation-maximization of a Gaussian mixtures}
\newacronym{cmr}{CMR}{cyclic mechanical reinforcement}
\newacronym{rgd}{RGD}{Arg-Gly-Asp}
\newacronym{ilk}{ILK}{integrin-linked kinase}
\newacronym{pinch}{PINCH}{particularly interesting new cysteine-histidine rich protein}
\newacronym{vasp}{VASP}{vasodilator-stimulated phosphoprotein}
\newacronym[\glslongpluralkey={integrin adhesion complexes}, \glsshortpluralkey={IACs}]{iac}{IAC}{integrin adhesion complex}
\newacronym[\glslongpluralkey={guanine nucleotide exchange factors}, \glsshortpluralkey={GEFs}]{gef}{GEF}{guanine nucleotide exchange factor}
\newacronym[\glslongpluralkey={GTPase activating proteins}, \glsshortpluralkey={GAPs}]{gap}{GAP}{GTPase activating protein}
\newacronym[\glslongpluralkey={kidney ankyrin repeat-containing proteins}, \glsshortpluralkey={KANK}]{kank}{KANK}{kidney ankyrin repeat-containing protein}
\newacronym[\glslongpluralkey={intercellular adhesion molecules}, \glsshortpluralkey={ICAMs}]{icam}{ICAM}{intercellular adhesion molecule}

\begin{document}
\pagenumbering{arabic} 
\onecolumn

\maketitle


\begin{abstract}

\section{Abstract}
Nascent adhesions are submicron transient structures promoting the
early adhesion of cells to the extracellular matrix. Nascent adhesions
typically consist of several tens of integrins, and serve as platforms
for the recruitment and activation of proteins to build mature focal
adhesions. They are also associated with early stage signalling and the
mechanoresponse. Despite their crucial role in sampling the local extracellular
matrix, very little is known about the mechanism of their formation.
Consequently, there is a strong scientific activity focused on
elucidating the physical and biochemical foundation of their
development and function. Precisely the results of this effort will be
summarized in this article.

\end{abstract}

\section{Introduction}
Integrin-mediated adhesion of cells and the associated mechanosensing is of 
monumental importance for the physiology of nearly any cell type~\citep{Kechagia2019}.
It often proceeds through the maturation of \glspl{na}, which are transient supramolecular 
assemblies, to \glspl{fa}, connecting a cell to the extracellular matrix or
another cell. \glspl{na} 
typically contain around 50~integrins~\citep{Changede2015}, and show a high turnover 
rate, with lifetimes of a bit over a minute~\citep{Choi2008}. 
\glspl{fa}, which arise upon the maturation of \glspl{na} by recruitment of 
numerous proteins to their cytoplasmic tails, form multimolecular integrin 
adhesion complexes. \Glspl{fa} are establishing the linkage between the \gls{ecm} 
and the actin cytoskeleton~\citep{Winograd-Katz2014}. However, another cytoskeletal element,
microtubules, also play an important role in adhesion and regulate the turnover
of adhesion sites~\citep{Bouchet2016, Chen2018}.

While the \glspl{fa} have been studied extensively in the last 
decades~\citep{geiger2009, Parsons2010, Wehrle-Haller2012, Geiger2011, Cooper2019, Green2019}, 
the smaller characteristic size of $\sim \SI{100}{\nano\meter}$ diameter make \glspl{na} 
significantly more elusive~\citep{Changede2015}. Studies of \glspl{na} 
require \gls{smlm} techniques to image below the diffraction limit of conventional 
light microscopy, and, furthermore, need to account for the short characteristic 
lifetime of few minutes~\citep{Changede2015}. The resulting scarcity of data 
associated with \glspl{na} makes the theoretical modeling very difficult.

In our current understanding, the \gls{na} formation follows three major steps~\citep{Sun2014}.
First, integrins activate, going from a state of low to high affinity through a 
conformational change. This can be induced by binding of activating proteins like 
talin or kindlin~\citep{Ye2013, Cluzel2005, Saltel2009, Changede2015, Ellis2014, Humphries2007}, 
or by binding to a ligand~\citep{Barczyk2010}. In the second step, the integrins
cluster into \glspl{na}, which show similar structures on substrates of different 
rigidities~\citep{Changede2015}, and are not reliant on \gls{m2} 
activity~\citep{Choi2008, Oakes2646, Bachir2014}. 
Finally, these clusters are either  disassembled, or they mature into \glspl{fa} 
and possibly further into fibrillar adhesions. 

Despite the efforts leading to our current understanding, the determinants of \gls{na} 
formation, turnover, or maturation, as well as their role in mechanosensing and 
signaling, are far from being fully resolved. However, the past two decades 
witnessed the emergence 
of several novel optical imaging techniques, technological advances in protein 
engineering and mass spectrometry analysis, as well as the expansion of 
theoretical modeling that now allow the investigation 
of protein organization of \glspl{na} at the nanoscale. Motivated by these perspectives, 
we here attempted to recapture recent advances in the field, while identifying open 
questions which we believe will be addressed in future research.

\section{Integrins as the key players in Nascent Adhesion}

\subsection{Integrin Physiology}

As their name suggests, integrins are cell adhesion proteins that are integral for 
numerous physiological functions. Examples include cell gene expression and 
differentiation, cell migration during embryonic development, immune response, 
or wound healing.
They sense the mechanical properties of the cell environment and provide signals 
necessary for cell survival, proliferation, and 
migration~\citep{Cooper2019, Green2019, Horton2015, Michael2020, Humphries2019}. 
Moreover, integrins are a prominent target for 
medication~\citep{Alday-Parejo2019, Bachmann2019}.

In humans, this broad range of functionalities is maintained by 
$24$~integrins~\citep{Shimaoka2003} built from $18~\alpha$- and $8~\beta$-subunits. 
Several combinations of $\alpha$ and $\beta$ subunits are possible, as either
is not necessarily limited to a single partnered subunit to form a 
heterodimer~\citep{Campbell2011, Hynes2002}.
They structurally form a headpiece and two legs~\citep{Xiong2001}. 
However, only integrins assembled as heterodimers in the endoplasmic reticulum are expressed 
on the cell surface, the composition of which cannot be reliably predicted by 
the mRNA expression levels~\citep{Hynes2002} of integrin subunits.
Integrin expression can be regulated by modulating their internalization and 
recycling, which contributes to the dynamic remodeling of adhesion~\citep{Moreno-Layseca2019}.

As integrins bind to more than one ligand, 
the 24 heterodimers are broadly categorized by their ligand specificity into 
\begin{enumerate*}[label=(\roman*)] \item \gls{rgd}
receptors, binding to fibronectin, fibrinogen, and thrombospondin, \item laminin 
receptors, \item collagen receptors, and finally \item leukocyte-specific receptors 
binding to different cell surface receptors such as \gls{icam}
and some extracellular matrix proteins~\citep{Humphries2006, Barczyk2010, Takada2007}
\end{enumerate*}. Additional ligands 
relevant in the immunological context are the intercellular adhesion molecules, 
immunoglobulin superfamily members present on inflamed endothelium, and antigen-presenting 
cells.
Besides binding to a wealth of ligands, the specificity for these interactions 
is promiscuous, as one integrin binds multiple ligands. Furthermore, it is also 
extremely redundant, as different integrins bind to the same ligand~\citep{Humphries2006}. 
These properties of integrins are essential for the interaction with the extracellular 
matrix and the consequent mechanotransduction. However, they make integrin research 
challenging at the cellular level.

Integrin binding and clustering provokes the formation of multimolecular \glspl{iac}
recruited to their cytoplasmic tails. The composition 
of \glspl{iac}, termed adhesome, have been analysed from cells seeded on fibronectin, 
using different methods~\citep{Byron2015, Jones2015, Kuo2011, Schiller2011, Zaidel-Bar2007}.  
This led to the definition of a fibronectin-induced meta adhesome composed of over 
2400 proteins which was further reduced to 60 core proteins, termed
consensus adhesome~\citep{Horton2015}.
Since integrins have no actin binding sites, they rely on a range of so-called 
adaptor proteins which bind to their cytoplasmic tails of integrins and bridge 
to the cytoskeleton.
There are four potential axes that link integrins to actin, 
namely
\begin{enumerate*}[label=(\roman*)]
\item \gls{ilk}-\gls{pinch}-1-kindlin,
\item \gls{fak}-paxillin,
\item talin-vinculin and 
\item $\alpha$-actinin-zyxin-\gls{vasp}
\end{enumerate*}~\citep{Horton2016, Horton2015, Humphries2019, Winograd-Katz2014}. 
The consensus adhesome also contains 
signalling molecules such as kinases, phosphatases, guanine 
nucleotide exchange factors (GEFs), GTPase activating proteins (GAPs) and 
GTPases~\citep{Horton2016}. Talin, a well-known activator discussed below, also
coordinates the microtubule cytoskeleton at adhesion sites through the interaction with
KN motif and \glspl{kank}~\citep{Bouchet2016, Chen2018, Paradzik2020, Sun2016},
which was shown to stimulate \gls{fa} turnover~\citep{Stehbens2012}.

A lot of detail regarding integrin structure and interactions with adaptor protein is 
obtained from molecular dynamics simulations, which starting with the seminal works on 
conformational changes in activation~\citep{Puklin-Faucher2006}, addressed integrin 
unfolding~\citep{Chen2011}, differences in integrin \glspl{tmd}~\citep{Pagani2018}, 
talin-integrin interactions also regarding the surrounding lipids~\citep{Kalli2017}, 
and interactions with other proteins~\citep{Shams2017}.

Integrin involvement in pathological conditions is mostly the consequence of 
changes in the expression,
either up- or down-regulation.
Prominent examples here are tumorigenesis but also the response to chemo- or 
radiotherapy~\citep{Cooper2019}. Therefore, integrin repertoire changes are an active target
for drug development in tumours with potential to inhibit metastasis, as well as 
to overcame resistance
to chemotherapy or radiotherapy. However, despite convincing experimental evidence 
that demonstrates the capacity of integrin inhibitors and monoclonal antibodies 
to contribute to inhibition of cancer progression, metastasis, or boost
therapeutic effects, no integrin-targeting drugs have been registered as anti-cancer 
drug~\citep{Alday-Parejo2019, Cooper2019, Desgrosellier2010, Dickreuter2017, Hamidi2018, Seguin2015}. 
Integrins are, nonetheless, used as targets in the prevention of blood clots during 
the opening of blood vessels in the heart~\citep{Tam1998}, multiple 
sclerosis~\citep{Polman2006} and Crohn’s disease~\citep{Gordon2001, Rosario2017}. 
Furthermore, since the accumulation of disorganized \gls{ecm} is modulated by several 
integrin heterodimers via activation of latent \gls{tgf}-$\beta$, the selected 
integrins are considered as promising therapeutic targets for fibrosis~\citep{Kim2018}.
Besides integrin up- or down-regulation, integrin mutations are also associated 
with some diseases like
junctional epidermolysis bullosa, caused by mutations in either integrin subunit of 
integrin $\alpha6\beta4$ forming \glspl{hd} or integrin $\alpha3$, which pairs with $\beta1$,
forming \glspl{fa}~\citep{McGrath2015, Walko2015}.
Furthermore, integrin related diseases may also be caused by impaired activation
as observed on platelets and leukocytes~\citep{Alon2003}. Integrins are also 
involved in bacterial~\citep{Hoffmann2011} and viral infections, either in 
attachment or internalisation~\citep{Hussein2015}, thus representing possible 
target molecules to combat infectious diseases.

Most integrin-related research, nevertheless, involves studies of mature adhesions. 
The physiological role of \glspl{na} is thus typically discussed in the context 
of the physiology of these superstructures. However, with the recently initiated 
debate that \glspl{na} may themselves act as signaling platforms, new perspectives 
in targeting \gls{na}-associated processes emerges. However, harnessing these 
possibilities  requires detailed knowledge of the sensory role of \glspl{na}, 
their dynamic behavior, and their regulation, which are all still poorly understood. 

\subsection{Integrin Activation}

Activation is the first step in the formation of \glspl{na} and is associated 
both with an integrin affinity change and the binding of integrins to extracellular 
ligands~\citep{Calderwood2004}. Activation as a term is also used to 
signify the switch to the \gls{eo} conformation, which is with 
the \gls{bc} and the \gls{ec} states, one of three major integrin 
conformations~\citep{Luo2007}.  The changes of conformation may be introduced by 
thermodynamic fluctuations~\citep{Sun2019}, but the switch is often induced by the 
very association of integrins with  ligands, adaptor proteins or Mn$^{2+}$.
Generally, each conformation has a specific 
affinity for ligands~\citep{Wang2799}, although all three conformations may be 
specific to one or more ligands.
Often though, the activated \gls{eo} state is 
the one with the highest binding affinity~\citep{Li2017}.

For example, prior to activation, the \gls{bc} state is the most common conformation 
of $\alpha5\beta1$ in the K562 chronic myelogenous leukemia cell line, 
making up for around $\text{99.76}\%$ of 
the population~\citep{Li2017}. Simultaneously, the \gls{ec} and \gls{eo} states 
contribute with  $0.09\%$ and $0.15\%$, respectively. However, $\alpha5\beta1$  
and $\alpha4\beta1$ in the \gls{eo} state have a 4000--6000~fold  and a 600--800~fold 
higher affinity for a ligand  compared to the \gls{bc} state~\citep{Li2018}. 
Notably, these affinities are measured for ligands in solution, where 
they do not induce integrin clustering~\citep{Cluzel2005}. 

In the environment 
of the plasma membrane, however, these affinities may change considerably due to 
the coupling to the membrane~\citep{Bell1978, Dembo1988, Bihr2012, Fenz2017}. 
Namely, the membrane, by its elasticity and fluctuations can change significantly 
the affinity of a bond, and can induce switches from low to high affinity 
states, even without changing the actual conformation of the proteins binding the ligands~\citep{Fenz2011, Kim2020}.
This mechanism of regulation of affinity was originally 
suggested by theoretical modeling~\citep{Bihr2012}, and was demonstrated for a 
variety of membrane associated ligand-receptor pairs~\citep{Bihr2015, Fenz2017}. 
However, its relevance for integrin binding remains to be shown explicitly, 
although it should be particularly relevant for the formation of \glspl{na}.   
Preliminary hints for the role of these mechanism come from mimetic liposome 
or bilayer model systems.  In the absence of adaptor proteins, activation of integrins 
was here successfully achieved by Mn$^{2+}$~\citep{Goennenwein2003, streicher2009}. 
The later was necessary to enable 
binding to \gls{rgd}~\citep{Smith2008}, which was however, very sensitive.  Mn$^{2+}$, 
however, may induce integrin conformations that can be different from a physiological 
ones~\citep{Ye2012}, which may contribute to the low binding yield, but cannot account 
for the observed variability of the data. 

In cellular systems, the activation may be induced by the binding to 
ligands~\citep{Park2016, Barczyk2010} but also by adaptor proteins~\citep{Calderwood2013, Ye2010}, 
for example, by talin~\citep{Cluzel2005, Saltel2009, Park2016}. Actually, already \gls{thd} 
was found to be enough to activate integrins~\citep{Calderwood1999}. Furthermore, \gls{thd} 
was found to synergize with the kindlin in integrin activation~\citep{Calderwood2013, Bledzka2012, Ma2008}, 
which in combination may promote binding to multivalent ligands~\citep{Ye2013}. 
However, there seems to be a competition between talin and kindlin, as the overexpression 
of kindlin-1 and kindlin-2 can both enhance and reduce integrin activation by \gls{thd}, 
depending on the integrin type~\citep{Harburger2009}. In other cases, kindlin 
overexpression showed only a small effect compared to 
\gls{thd}~\citep{Ye2010, Ye2012, Ye2013, Ma2008, Shi2007}.

\begin{figure}[h!]
\begin{center}
\includegraphics[width=15cm]{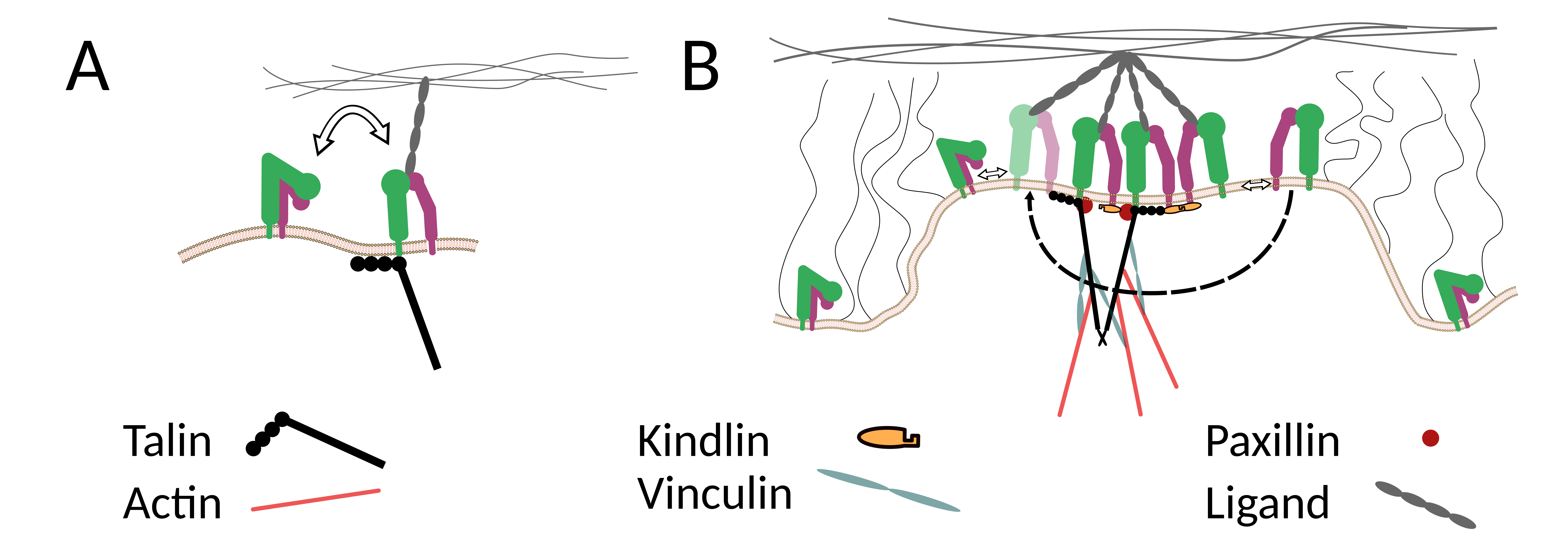}
\end{center}
\caption{Current picture of integrin activation and clustering. (\textbf{A}) Fluctuations
	between the closed and open integrin activation state are stabilized by binding
	of intracellular talin, or extracellular ligands. Although kindlin also plays a 
	role, talin shows the strongest effect in stabilizing the open state in early 
	adhesions. (\textbf{B}) Initial liganded integrins create a region with increased
	binding probability, by membrane and glycocalyx deformation. Clustering is further 
	amplified by multivalent ligands, dimerizing adaptor proteins, established scaffolding
	structures and \gls{tmd} interactions.}\label{fig:1}
\end{figure}

Binding of ligands allows for anchoring of integrins and the exertion of forces 
by the actomyosin network, but also by the fluctuations of the 
membrane (Figure~\ref{fig:1}A). The consequence is a dynamic change of the free energy
of \gls{bc}, \gls{ec} and \gls{eo} states. The importance of this effect was 
highlighted in the model of \citeauthor{Li2017a}~\citep{Li2017a}, who find a 
sigmoidal dependence of activation probability with respect to the applied force,
where activation here signifies binding of both ligand and adaptor protein. 
This resulted in full activation of all integrins over few \si{\pico\newton} for 
a wide range of adaptor protein concentrations, permitting a quick response to 
mechanical stimuli. Comparable behaviour was found in fibroblasts, which reinforce 
early integrin adhesions $\leq \SI{5}{\second}$ under load by binding additional 
integrins~\citep{Strohmeyer2017}.

The response of integrins to force has to be considered also in the context of 
the catch-bond effect. Namely, unlike slip bonds, which subject to force 
show an increase in the unbinding rate~\citep{Bell1978}, catch bonds are stablized 
by force~\citep{Dembo1988}. So far, in the case of integrins, both behaviors were 
found for different force regimes, which lead to the introduction of the term catch-slip 
bond. The latter was observed for example for $\alpha_5\beta_1$~\citep{Kong2009}, 
$\alpha_4\beta_1$~\citep{Choi2014}, and $\alpha_L\beta_2$~\citep{Chen2010}. In 
the case of $\alpha_5\beta_1$, catch bond formation seemed to involve 
the headpiece, but not integrin extension~\citep{Kong2009}.

Other distinct mechanisms that, similarly to catch bonds, strengthen integrin attachments 
in a force dependent manner are \gls{cmr}~\citep{Kong2013}, 
and the so called dynamic catch~\citep{Fiore2014}. In \gls{cmr}, an increase of 
bond lifetime occurs over several loading-unloading cycles. In dynamic 
catch, the force response is regulated synergistically by the of binding of an 
additional co-receptor to form a trimolecular complex with the integrin and the 
common ligand. In mimetic, actin free systems, cyclic application of force also 
resulted in bond-strengthening~\citep{Smith2008}, which was shown to emerge from a 
thermodynamic response of the ensemble. Application of a pulling force~\citep{smith2005b} 
induced a regrouping of bonds from sparse 
configurations to clusters in which cooperative response is 
allowed (strengthening each bond in average), and a new 
thermodynamic state is established~\citep{Smith2008}.  Interestingly, recent 
work showed that even in after the links with the cytoskeleton are fully 
established, most integrins existed in the state of near-mechanical equilibrium~\citep{Tan2020}.

This complex behavior of integrins is cast into two major activation models, 
the so-called inside-out, induced by cytoplasmic factors, and the outside-in, 
where the activation results from binding to extracellular ligands~\citep{Wang2799}. 
However, in the formation of \glspl{na}, the time-sequence of binding events is not yet 
fully established, and it is not clear to what extend the cell relies on either method
in order to form such complex structures as \glspl{na}.

\section{Establishment of Nascent Adhesions and Larger Structures}
\subsection{Formation of Integrin Clusters}

Clustering of integrins (Figure~\ref{fig:1}B), with and without the help of adaptor proteins 
and independent of F-actin~\citep{Cluzel2005} and \gls{m2} activity~\citep{Choi2008}, 
builds the second step of \gls{na} formation. Understanding of this process is greatly 
facilitated by the emergence of \gls{sr} techniques. The latter provide optical images 
with spatial resolutions below the diffraction limit of light of the order of $\sim \SI{100}{\nano\meter}$~\citep{Sigal2018}. 
Therefore, it should be possible to resolve the dynamic nanoscale organization of \glspl{na} 
and the force transduction across individual components within \glspl{fa}.  However, 
quantitative investigations of \glspl{na}, are still lacking. The main reason is that 
existing \gls{smlm} techniques require cluster 
analysis tools, which have been developed for relatively simple cases, such as membrane 
protein clusters without strong heterogeneity in size, shape, and density~\citep{Nicovich2017, Nieves2020}. 
Several studies have addressed this  by designing novel approaches to investigate the 
inner architecture of \glspl{na} and \glspl{fa}, such as one based on the \gls{emgm}~\citep{Deschout2017}. 

So far, however, various nanoscale distributions have been observed for integrins. 
Clusters as small as \numrange[range-phrase = --]{2}{3} integrins were reported using 
electron microscopy~\citep{Li2003}, while clusters observed in \gls{smlm} range from 
tens to hundreds of molecules. Some of the first application of SR techniques yielded \SI{100}{\nano\meter} 
large \glspl{na}, containing on average 50 integrins~\citep{Changede2015}. This data is 
contrasted by a more recent work with improved \gls{emgm} method used on the \gls{palm}
data, when it was determined that \glspl{fa} cover 
areas between \numrange[range-phrase = ~and~]{0.01}{1} \si{\micro\meter\squared}. 
Using \gls{emgm}, localization uncertainties, an important and unavoidable aspect of 
any \gls{smlm} experiment, could be corrected showing that the assemblies 
contained \numrange[range-phrase = ~to~]{10}{100} localizations, and exhibited strong 
eccentricities~\citep{Deschout2017}. Notably, most existing \gls{smlm} clustering methods 
ignored this effect, which can lead to substantial overestimation of the size of 
identified localization structures.  

While the dynamic behavior of \glspl{na} is still an open problem, it is nevertheless 
clear that clusters allow for quick rebinding after bond failure~\citep{Bihr2012, Sun2019}, 
and the control over maturation or disassembly~\citep{Schmidt2015}. Furthermore, clusters 
could serve as platforms for rigidity sensing~\citep{Wolfenson2015}, however, it is still 
unclear which point in the process of \gls{na} assembly corresponds to the onset of signaling.  

In the absence of detailed microscopy studies, even the necessary conditions for the formation 
of these meta-stable aggregates are unclear. Some studies report that integrin activation 
is indispensable for clustering~\citep{Cluzel2005}, promoting the nucleation of new 
structures~\citep{Saltel2009}. These results are contrasted by experimental findings that show 
both active and inactive integrin nanoclusters in \glspl{fa}~\citep{Spiess1929}, obtained 
using extended state specific antibodies that co-localized with talin, kindlin-2 and vinculin. 
The existence of inactive clusters could suggest the affinity for ligands in the inactive 
states is sufficiently large to promote nucleation of domains, although with smaller 
probability than in the active state. Alternatively, one could conclude that ligand binding 
is not necessary for clustering, although it is possible 
that ligand bound states preceded cluster formation.

Further scenarios suggest a link between integrin activation and clustering mediated by 
lateral interactions between tails of \gls{tmd}~\citep{Ye2014, Mehrbod2013, Li2003}. 
However, limited size of \glspl{na}~\citep{Changede2015, Changede2017} requires further 
regulation of such interactions. Moreover, the necessary activation energy between 
\glspl{tmd} also seems too high to overcome without help~\citep{Mehrbod2013}. In simulations, 
fewer integrins cluster when the lateral integrin interactions are weak~\citep{Bidone2019}.
In addition, the \gls{tmd} could not 
drive the clustering in Mn$^{2+}$ activated integrins, without ligands present~\citep{Cluzel2005}.  

With mobile ligands, on the other hand, Mn$^{2+}$ activated integrins formed small adhesion domains, 
which significantly increased in size if integrins themselves were maintaining lateral 
mobility prior to the establishment of bonds~\citep{Smith2008}. In this case the clustering 
of bound integrins was mediated by the deformed membrane. The nature and magnitude of these 
forces could be clearly elucidated~\citep{Janes2019}, and were proposed to play an important 
role in the cluster nucleation and growth~\citep{Bihr2012}. Given that these types of 
interactions are not protein specific, they should also be seen in other binding systems. 
Indeed, correlations in membrane dynamics and topography with cell spreading reported recently 
in several studies of cell adhesions~\citep{pierres2008, Perez2008, LamHui2012}, and 
systematically in reconstituted passive systems based on \glspl{guv}~\citep{smith2009}, 
including those involving integrins~\citep{Goennenwein2003, streicher2009, Smith2008}. However, 
this mechanism remains to be directly confirmed for integrins in the cellular context.

\begin{figure}[h!]
\begin{center}
\includegraphics[width=15cm]{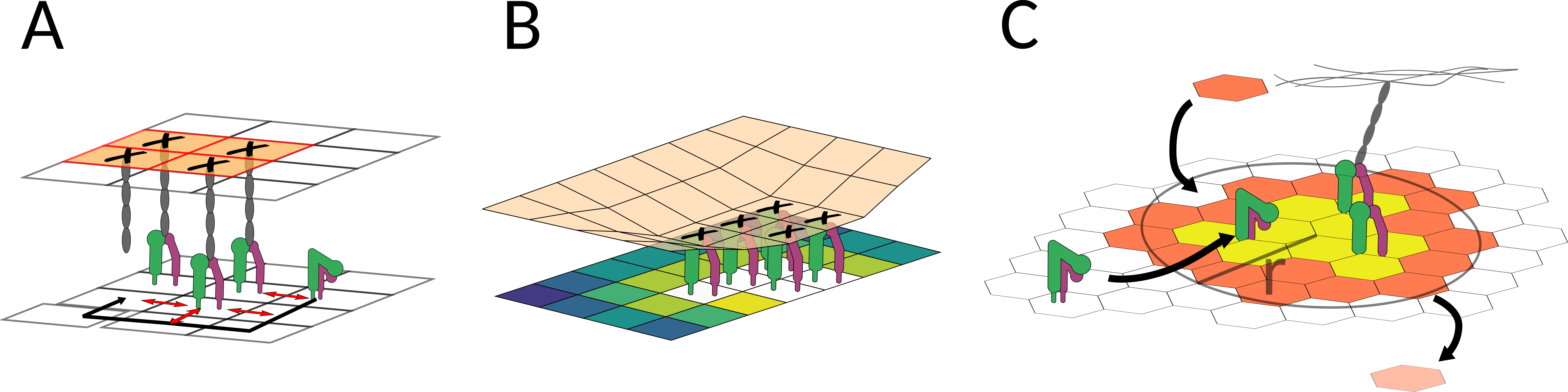}
\end{center}
\caption{Different theoretical models of \gls{na} formation, or parts thereof. 
	(\textbf{A}) \gls{tmd} interactions between neighbouring integrins as well as
	clustered ligands can cooperatively increase clustering further by guiding
	close-by integrins to free ligand binding sites~\citep{Brinkerhoff2005}. 
	(\textbf{B}) Membrane and
	glycocalyx deformations can create regions with enhanced binding probabilities,
	catching diffusing free integrins~\citep{Fenz2017}. 
	(\textbf{C}) In a rate equation model of
	the \gls{na} formation, the \gls{na} size is defined only by adaptor protein
	concentration, which are stabilized through present integrins, by reducing their 
	unbinding rates~\citep{MacKay2019}. The honeycomb grid as 
	well as the hexagonal adaptor proteins shape and size are for visualization only.}\label{fig:2}
\end{figure}

Most of membrane-related mechanisms include implicitly the existence of the cellular 
glycocalyx~\citep{bruinsma2000, smith2009}, which was indeed found to play an important 
role in integrin clustering~\citep{Paszek2014}. The compression and consequent expulsion 
of glycocalyx, by the formation of the initial bond primes the surroundings for further 
interactions. Concomitantly the membrane deforms towards the 
ligand~\citep{Janes2019}, creating a microenvironment in which the additional bonds have a 
much higher likelihood to form~\citep{Bihr2015, Fenz2017} (figure~\ref{fig:2}B). The 
tension on the bond furthermore increases their lifetimes~\citep{Kong2009}, in a 
synergistic fashion. These effects can be further strengthened by membrane 
thermal~\citep{helfrich1978} and active fluctuations~\citep{Turlier2018}, which adds 
to the portfolio of forces acting on \glspl{na}, the latter being regulators of 
adhesion formation~\citep{Strohmeyer2017, Li2017a, Oakes2646}.

While the interplay between these many factors contributing to the \gls{na} formation 
in its early stages is not yet fully understood, there is a consensus that integrin 
activation increases binding to anchored, clustered ligands. A strong 
increase in the number of spreading cells was found for a basic pattern of 4 ligands 
at $\sim \SI{60}{\nano\meter}$ distance compared to 3 ligands at the  same density~\citep{Schvartzman2011}. 
Similar effects are well captured by theoretical works. Simulations 
can also account for the interplay between different integrin types, as demonstrated on the example 
the competition between ligand binding and clustering of $\beta_1$ and $\beta_3$~\citep{Bidone2019}. Similarly, the 
effect of closely spaced multivalent ligands was also captured in simulations in which clusters of 
more than two integrins can form through dimerization, if the interactions are weak~\citep{Brinkerhoff2005}, 
allowing transient dimer interactions with switching partners. A similar result was confirmed by an 
agent-based model~\citep{Jamali2013}, where large agglomerates of ligands provide largest integrin clusters.

Besides ligands, a number of adaptor proteins have been involved in cluster formation. The 
most prominent examples are kindlin~\citep{Ye2013, Changede2015} and talin (particularly 
its head domain)~\citep{Cluzel2005, Saltel2009, Changede2015, Calderwood2013}, that have been 
already implicated in integrin activation~\citep{Theodosiou2016, Moser2008, Zhang2008, Ye2013}. 
Their association to integrins then of course promotes binding to ligands and clustering. Furthermore, 
both kindlin~\citep{Li9349, Kammerer9234} and talin~\citep{Golji2014} have a capacity for dimerization. 
For example, talin rod, which, using its integrin binding site, can rescue clustering in talin depleted 
cells~\citep{Changede2015}. However, the efficiency of talin rod fragments was found to be inferior 
to the full length talin~\citep{Saltel2009}. This points to a possible synergy between dimerization 
and binding the monovalent, and even more so multivalent ligands (figure~\ref{fig:2}A). 

These potentially complex stoichiometries are, however, a challenge for \gls{smlm}. Complex 
photo-physics of interacting fluorophores can lead to over-counting of molecules at a given 
location~\citep{Annibale2011}. This complicates the accurate determination of protein 
stoichiometries from the data. It is, however, possible to estimate the number of 
labelled-proteins contained in a single cluster~\citep{Annibale2011a, Baumgart2016, Spahn2016, Pike2019}, 
but more accurate quantifications are needed and their accuracy demands further validation. 
A recently developed supervised machine-learning approach to cluster analysis can be an 
interesting candidate to cope effectively with \glspl{na} sample heterogeneity~\citep{Williamson2020}. 
It was successfully applied on data of the C-terminal Src kinase and the adaptor PAG in 
primary human T cell immunological synapses~\citep{Williamson2020}, but was not yet tested 
for talin-integrin complexation or even more generally on \gls{na} data. 

Talin rod has an additional property important for the formation of \glspl{na}, namely 
it possesses a binding site for vinculin. Vinculin is a cytoskeletal protein with binding 
sites, besides talin, for actin, $\alpha$-actinin, and lipids and it is usually 
associated with the force transmission. However, recently it was found that the talin rod domain 
is available to vinculin in a force-independent manner upon the release of talin 
autoinhibition~\citep{Dedden2019, Atherton2020}. This would suggest that vinculin could 
play a role in \glspl{na} even before the cytoskeletal forces are involved.

The integration of vinculin could be facilitated by PI(4,5)P$_2$. This phospholipid 
regulates talin-integrin interactions at the level of the membrane. Its association with 
the $\beta_3$ units opens a binding site for the integrin on the talin head, hence controlling 
the talin auto-inhibition. Furthermore, the PI(4,5)P$_2$ interaction with the integrin creates 
a salt bridge toward the membrane that prevents the close interactions of 
the $\alpha$ and $\beta$ subunits. Therefore, the integrin remains in an activated, 
clustering-competent state~\citep{Saltel2009, Dedden2019, Cluzel2005}. Consistently with 
these findings, sequestering of PI(4,5)P$_2$ diminishes the formation of clusters~\citep{Cluzel2005}.

One more protein strongly investigated in the context of integrin-clustering is $\alpha$-actinin~\citep{Sun2014}. 
It is microfilament protein necessary for the attachment of actin. Both positive and negative 
effects were demonstrated for $\beta_1$ and $\beta_3$ integrins, respectively~\citep{Shams2017, Roca-cusachs2013},
which could relate to an integrin crosstalk strategy~\citep{Bharadwaj2017}. However, 
its role in clustering is still debated~\citep{Theodosiou2016}. 

This large number of molecular players involved through various interactions poses a significant 
challenge for comprehending the formation of \glspl{na}. A promising approach that can address 
this diversity is theoretical modeling based on rate equations~\citep{Schwarz2006, Harland2011, Walcott2010, Li2010, Zhu2000}. 
An early attempt focused purely on the talin-PIP2 interaction and is therefore limited in scope~\citep{Welf2012}. 
However, more recently, emulating the formation of entire \glspl{na} has been attempted~\citep{MacKay2019}. 
In the latter case, a higher emphasis is set on the crosslinking function of adaptor proteins (figure~\ref{fig:2}C). 
The model reproduces some features found in experiments, for example the limited area of \gls{na}, 
even at high integrin density. It also predicts  unliganded clusters. Actually, the model 
defines \glspl{na} as plaques of adaptor proteins and not as integrin clusters. Integrins 
stabilize the plaque but are not required, which allows for the possibility of preclustering adaptor 
proteins in the absence of integrins. 

\subsection{Cluster disassembly or maturation}

Clustering of \glspl{na} into \glspl{fa} or disassembly is the last step
in the \gls{na} life time. The fate of \glspl{na} depends on the cell type, protein composition, and mechanical 
properties of the substrate~\citep{Parsons2010}, as well as the attachment to the 
actin cytoskeleton and both \gls{m2} isoforms~\citep{Vicente-Manzanares2007}. 

Most \glspl{na} disassemble when the lamellipodium moves past them~\citep{Choi2008} following 
one of several competing ways of \gls{na} disassembly, as discussed in the literature~\citep{Gardel2010a}. Particularly 
well studied is the role of the non-receptor tyrosine kinase \gls{fak} that is known to regulate 
adhesion disassembly~\citep{Webb2004}, possibly through talin proteolysis~\citep{Lawson2012a}. 
In addition, \gls{fak} might be inhibited at the leading edge of the lamellipodium through 
interactions with Arp2/3, a protein complex that is related to the actin branching~\citep{Swaminathan2016}. 
Because the regulation of rearrangements of the actin cytoskeleton is crucial for filopodia 
extension~\citep{He2017} and lamellipodia formation~\citep{Small2002}, \gls{fak} is implicated 
in the spatial control of the advance of the leading edge and the \gls{na} disassembly. This 
regulation is facilitated by the binding of \gls{fak} to paxillin, which is also recruited 
by kindlin~\citep{Bottcher2017,Zhu2019, Humphries2007}, to further control the  adhesion 
turnover~\citep{Choi2011, Shan2009}. Interestingly enough, vinculin, can impede the \gls{fak}-paxillin 
interaction~\citep{Subauste2004}, while \gls{fak} may play a role in recruiting talin 
to \gls{na} sites~\citep{Lawson2012, Lawson2012a}, as well as Arp2/3~\citep{Lawson2012a}. 
As \gls{fak} also plays an important role in signalling~\citep{Swaminathan2016}, 
it is difficult to unravel the precise dynamical interactions in \glspl{na}.

Another simple way to dissolve \glspl{na} is by offering  soluble ligands~\citep{Cluzel2005}. 
The latter either exhibit a lateral pressure on the \gls{na} site or they compete for the integrin 
upon stochastic unbinding~\citep{smith2006a}. Since the 3D affinity of soluble ligands is always larger 
than the 2D affinity of surface-confined ligands, the cluster becomes unstable.

Alternative to disassembling is the sequential maturation of \glspl{na} into \glspl{fa}, 
and in some cases, to centrally positioned 
fibrillar adhesions enriched in tensin and integrin $\alpha5\beta1$~\citep{Iwamoto2015}. The 
inhomogeneous structure of these assemblies were observed already over a decade ago 
using \gls{palm} for imaging \gls{fa} proteins~\citep{betzig2006}. Just one year 
later, \citeauthor{Shroff2007}~\citep{Shroff2007} used dual-color \gls{palm} to determine the ultrastructural relationship 
between different pairs of \gls{fa} proteins. The consensus today is that integrins 
bind via talin and other adaptor proteins with the actin cytoskeleton, allowing \gls{m2}
generated forces to act on the clusters~\citep{Yu2011}. Under force the bonds strengthen, and tilt in 
the direction of the retrograde actin flow~\citep{Nordenfelt2017}. The reinforcement of integrin assemblies 
is further promoted by the recruitment of vinculin~\citep{Huang2017}, the crosslinking by myosins~\citep{Burridge2016},
and the exposure of  force-dependent cryptic binding sites~\citep{Ciobanasu2014, Yao2015, Yao2016} 
that allow for the attachment of other adhesome proteins. Finally, through mature adhesions, 
force is propagated from the \gls{ecm} to the actin cytoskeleton over the unfolding talin that permits 
vinculin binding~\citep{Asaro2019}, resulting in a strong signalling cascade and mechanoresponse 
along the adhesome, that regulates a number of physiological processes in cells.

\section{Challenges and Perspectives in the Field of Nascent Adhesions}

As presented in the above discussion, many molecular players contributing to the formation of \glspl{na} 
have been identified, and their mutual interplay have been established, although further investigation of 
specific interactions is necessary. For example, very little is known about the crosstalk between 
integrins of the same and different types, recently reported for early 
adhesions~\citep{Bharadwaj2017, Diaz2020, Strohmeyer2017}. However, new research avenues in studying 
molecular interactions in \glspl{na} can emerge only from advances in the development of robust 
quantitative colocalization analysis~\citep{Levet2019}. This needs to be accompanied by progress 
in genome editing and novel protein labelling strategies that could enable quantitative \gls{sr}. 
So far, only very few colocalizations of active integrins with talin and kindlin~\citep{Spiess1929} 
and vinculin and talin~\citep{Xu2018} could be observed. This can be either a technical issue, 
associated with protein expression,  probe photophysics, and the limited choice of labeling pairs 
and fluorophores. It could also point to unknown integrin regulators~\citep{Spiess1929}, and hidden 
interactions that remain to be revealed. For this purpose, molecular dynamics simulations will 
become increasingly important as they provide unmatched details in competing binding 
interactions~\citep{Mehrbod2013}. 
With appropriate level of coarse-graining larger complexes and 
slower structural changes are becoming within reach of \gls{md} simulations, which now can 
explicitly address integrin activation and clustering. 

Probably, however, the most acute issue is the spatiotemporal evolution of \glspl{na} 
and the role of complex stoichiometries. Namely, the dynamics of \glspl{na} is subject 
of intense debate as the constitutive clusters could be either stationary or showing 
stochastic transient immobilization~\citep{Spiess1929}. This problem is very closely related 
to the sensory capacity of \glspl{na} and the onset of signaling, that are equally ununderstood.
Resolution of these open questions requires new techniques that can deal with the fast 
molecular turnover within \glspl{na}. However, this is still a significant challenge for 
the single molecule localization microscopy~\citep{Orre2019} such as \gls{palm}~\citep{betzig2006},
\gls{storm}~\citep{rust2006} or 
\gls{sofi}~\citep{dertinger2009}). First promising insights into the dynamics of \glspl{na}, 
nevertheless were provided by the \gls{sptpalm}~\citep{Inavalli2019}, which revealed 
integrin cycling between free diffusion and immobilization, while transient interaction 
with talins promoted integrin activation and immobilization~\citep{Rossier2012}. Further 
studies of integrin dynamics will require techniques such that can operate at micro 
second-time scales with \SI{1}{\nano\meter} precision of molecules located few nanometers
apart. An example of such a method is minimal photon fluxes nanoscopy~\citep{Balzarotti2017}, 
although other approaches are starting to appear and will need to be employed in the 
research of \glspl{na}.  

Another challenge in studies of \glspl{na} is the impact of force. Although not strictly related 
to actomyosin activity, forces on integrin complexes arise due to the spacial confinement of the 
molecular players and result in load-dependent competition for binding partners. Different sources 
of forces may play a role in \gls{na} formation, prior to their maturation into \glspl{fa}. 
Specifically, the glycocalyx and the membrane are anticipated to generate relatively strong 
tensions and direct stochastic  forces on the individual integrins and the 
clusters~\citep{Paszek2014, Li2017a, Strohmeyer2017, Sengupta2018}. The understanding  of these 
effects relies on the development of force sensors~\citep{Tan2020}, and techniques which combine 
the force application with superresolution microscopy. Furthermore, given the intrinsically 
non-equilibrium and noisy setting, theoretical support in formulating and validating the 
appropriate hypothesis on the role of confinement is necessary. 

A particularly useful tool in the research of integrin adhesion so far have been functionalized 
substrates~\citep{Goennenwein2003, Liu2014, Schvartzman2011, Changede2019}. Manipulation with 
their stiffness, spatial coordination and mobility of binders allowed to provide mechanical 
cues which could be exploited to resolve the response of different cell models. Based on 
this long standing success, it is expected that patterned substrate will continue to play 
an important role in studies of \glspl{na}. Especially interesting should be their combination 
with specifically designed cell models that express different types of integrins on the 
plasma membrane surface. 

Furthermore, these substrates could be very successfully combined with reconstituted 
systems. The latter serve as an ideal  bridge between the biological complexity and 
theoretical modeling. Reconstituted systems, typically based on giant lipid vesicles, were 
instrumental in elucidating the role of mechanical properties of binders, as the role of 
the receptor and ligand density and mobility in the cell recognition process~\citep{smith2009}. 
Furthermore, vesicle-substrate adhesion was successfully used to study the physical 
mechanisms that regulate ligand-receptor binding, including the role of stochastic 
membrane deformations, fluctuations and composition, as well the steric repulsion role of 
the glycocalyx~\citep{Sengupta2018}. However, the simplicity of these assemblies may not 
represent the appropriate biological complexity of \glspl{na}. To circumvent that issue, 
more recently \glspl{dsguv} were designed that can be sequentially loaded with 
talin and kindlin~\citep{Weiss2017}. These systems show great potential for the studies 
of \glspl{na}, and could be used to drive the development and validation of theoretical 
models and simulations used to describe the growth process. 

At the current stage,  most theoretical models that attempt to capture the formation 
of \glspl{na} account for the  molecular complexity of the system, but capture the 
bio-mechanical context only implicitly, if at all. There is also another class of modes 
that is capable of resolving the stochastic nature of \gls{na} formation and the forces 
acting on the bonds with relatively high level of detail, but they are nearly void of 
molecular information. Future efforts are likely to bring closer these two distinct 
families of approaches, with the aim of providing a more reliable foundation that is 
required to capture the development of \glspl{na} in a predictive manner.  

Finally, it is a hope that the lessons learned in the studies of \glspl{na} may be 
useful in the context of other integrin-based structures. For example, in \glspl{hd}, 
linked to the internal keratin intermediate filament network, $\alpha6\beta4$ integrins 
mediate adhesion of epithelial cells to the underlying basement membrane~\citep{Walko2015}. 
In reticular adhesions, which serve to maintain the attachment of cells to the 
extracellular matrix during mitotic rounding and division, $\alpha V\!\beta5$ integrins, 
clathrin and endocytic adaptors also form adhesive 
complexes~\citep{Lock2018, Elkhatib2017, Grove2014, Leyton-Puig2017, Lock2019}. 
Currently, it is not 
known whether these different types of adhesion have precursor structures analogous 
to \glspl{na}. However, it is highly likely that tools, methods and approaches developed 
in studies of \glspl{na} may prove to be useful in these potentially different settings.

In closing, we strongly believe that a jointly advances in superresolution microscopy, 
the development of model systems and technology for manipulations of proteins, as well as 
theoretical approaches is required to further propel our understanding in molecular 
mechanisms of integrin organization, stoichiometry and dynamics at the nanoscale. This 
will not only allow us to rationalize the observed phenomena, but also gain important 
concepts and tools that can be used to resolve the physiological role of integrin based 
structures, but can be further applied beyond the \gls{na} research.

\section*{Conflict of Interest Statement}
The authors declare that the research was conducted in the absence of any commercial or financial relationships that could be construed as a potential conflict of interest.

\section*{Author Contributions}
HS conceived and wrote the first draft of the manuscript under the supervision of ASS. All authors contributed to the writing of the final paper. 

\section*{Funding}
ASS and HS thank the joint German Science Foundation and the French National Research Agency project SM 289/8-1, AOBJ: 652939, and the Excellence Cluster: Engineering of Advanced Materials at FAU Erlangen for support. AAR thanks the Croatian Science Foundation (Grant IP-2019-04-1577 to AA-R).
We also acknowledge support by the DFG Research Training Group 1962, Dynamic Interactions at Biological Membranes.  

\section*{Acknowledgments}
AR acknowledges the support of the Max Planck-EPFL Center for Molecular Nanoscience and Technology. 
\bibliographystyle{unsrtnat} 
\bibliography{bib}

\begin{thebibliography}{185}
\providecommand{\natexlab}[1]{#1}
\providecommand{\url}[1]{\texttt{#1}}
\expandafter\ifx\csname urlstyle\endcsname\relax
  \providecommand{\doi}[1]{doi: #1}\else
  \providecommand{\doi}{doi: \begingroup \urlstyle{rm}\Url}\fi

\bibitem[Kechagia et~al.(2019)Kechagia, Ivaska, and Roca-Cusachs]{Kechagia2019}
Jenny~Z. Kechagia, Johanna Ivaska, and Pere Roca-Cusachs.
\newblock {Integrins as biomechanical sensors of the microenvironment}.
\newblock \emph{Nature Reviews Molecular Cell Biology}, 20\penalty0
  (8):\penalty0 457--473, aug 2019.
\newblock ISSN 1471-0072.
\newblock \doi{10.1038/s41580-019-0134-2}.

\bibitem[Changede et~al.(2015)Changede, Xu, Margadant, and
  Sheetz]{Changede2015}
Rishita Changede, Xiaochun Xu, Felix Margadant, and Michael~P. Sheetz.
\newblock {Nascent Integrin Adhesions Form on All Matrix Rigidities after
  Integrin Activation}.
\newblock \emph{Developmental Cell}, 35\penalty0 (5):\penalty0 614--621, dec
  2015.
\newblock ISSN 15345807.
\newblock \doi{10.1016/j.devcel.2015.11.001}.

\bibitem[Choi et~al.(2008)Choi, Vicente-Manzanares, Zareno, Whitmore, Mogilner,
  and Horwitz]{Choi2008}
Colin~K. Choi, Miguel Vicente-Manzanares, Jessica Zareno, Leanna~A. Whitmore,
  Alex Mogilner, and Alan~Rick Horwitz.
\newblock {Actin and $\alpha$-actinin orchestrate the assembly and maturation
  of nascent adhesions in a myosin II motor-independent manner}.
\newblock \emph{Nature Cell Biology}, 10\penalty0 (9):\penalty0 1039--1050, sep
  2008.
\newblock ISSN 1465-7392.
\newblock \doi{10.1038/ncb1763}.

\bibitem[Winograd-Katz et~al.(2014)Winograd-Katz, F{\"{a}}ssler, Geiger, and
  Legate]{Winograd-Katz2014}
Sabina~E. Winograd-Katz, Reinhard F{\"{a}}ssler, Benjamin Geiger, and Kyle~R.
  Legate.
\newblock {The integrin adhesome: from genes and proteins to human disease}.
\newblock \emph{Nature Reviews Molecular Cell Biology}, 15\penalty0
  (4):\penalty0 273--288, apr 2014.
\newblock ISSN 1471-0072.
\newblock \doi{10.1038/nrm3769}.

\bibitem[Bouchet et~al.(2016)Bouchet, Gough, Ammon, van~de Willige, Post,
  Jacquemet, Altelaar, Heck, Goult, and Akhmanova]{Bouchet2016}
Benjamin~P Bouchet, Rosemarie~E Gough, York-Christoph Ammon, Dieudonn{\'{e}}e
  van~de Willige, Harm Post, Guillaume Jacquemet, AF~Maarten Altelaar,
  Albert~JR Heck, Benjamin~T Goult, and Anna Akhmanova.
\newblock {Talin-KANK1 interaction controls the recruitment of cortical
  microtubule stabilizing complexes to focal adhesions}.
\newblock \emph{eLife}, 5, jul 2016.
\newblock ISSN 2050-084X.
\newblock \doi{10.7554/eLife.18124}.

\bibitem[Chen et~al.(2018)Chen, Sun, and F{\"{a}}ssler]{Chen2018}
Nan-Peng Chen, Zhiqi Sun, and Reinhard F{\"{a}}ssler.
\newblock {The Kank family proteins in adhesion dynamics}.
\newblock \emph{Current Opinion in Cell Biology}, 54:\penalty0 130--136, oct
  2018.
\newblock ISSN 09550674.
\newblock \doi{10.1016/j.ceb.2018.05.015}.

\bibitem[Geiger et~al.(2009)Geiger, Spatz, and Bershadsky]{geiger2009}
Benjamin Geiger, Joachim~P. Spatz, and Alexander~D. Bershadsky.
\newblock {Environmental sensing through focal adhesions}.
\newblock \emph{Nature Reviews Molecular Cell Biology}, 10\penalty0
  (1):\penalty0 21--33, jan 2009.
\newblock ISSN 1471-0072.
\newblock \doi{10.1038/nrm2593}.

\bibitem[Parsons et~al.(2010)Parsons, Horwitz, and Schwartz]{Parsons2010}
J.~Thomas Parsons, Alan~Rick Horwitz, and Martin~A. Schwartz.
\newblock {Cell adhesion: integrating cytoskeletal dynamics and cellular
  tension}.
\newblock \emph{Nature Reviews Molecular Cell Biology}, 11\penalty0
  (9):\penalty0 633--643, sep 2010.
\newblock ISSN 1471-0072.
\newblock \doi{10.1038/nrm2957}.

\bibitem[Wehrle-Haller(2012)]{Wehrle-Haller2012}
Bernhard Wehrle-Haller.
\newblock {Structure and function of focal adhesions}.
\newblock \emph{Current Opinion in Cell Biology}, 24\penalty0 (1):\penalty0
  116--124, feb 2012.
\newblock ISSN 09550674.
\newblock \doi{10.1016/j.ceb.2011.11.001}.

\bibitem[Geiger and Yamada(2011)]{Geiger2011}
Benjamin Geiger and Kenneth~M. Yamada.
\newblock {Molecular Architecture and Function of Matrix Adhesions}.
\newblock \emph{Cold Spring Harbor Perspectives in Biology}, 3\penalty0
  (5):\penalty0 a005033--a005033, may 2011.
\newblock ISSN 1943-0264.
\newblock \doi{10.1101/cshperspect.a005033}.

\bibitem[Cooper and Giancotti(2019)]{Cooper2019}
Jonathan Cooper and Filippo~G. Giancotti.
\newblock {Integrin Signaling in Cancer: Mechanotransduction, Stemness,
  Epithelial Plasticity, and Therapeutic Resistance}.
\newblock \emph{Cancer Cell}, 35\penalty0 (3):\penalty0 347--367, mar 2019.
\newblock ISSN 15356108.
\newblock \doi{10.1016/j.ccell.2019.01.007}.

\bibitem[Green and Brown(2019)]{Green2019}
Hannah~J. Green and Nicholas~H. Brown.
\newblock {Integrin intracellular machinery in action}.
\newblock \emph{Experimental Cell Research}, 378\penalty0 (2):\penalty0
  226--231, may 2019.
\newblock ISSN 00144827.
\newblock \doi{10.1016/j.yexcr.2019.03.011}.

\bibitem[Sun et~al.(2014)Sun, Lambacher, and F{\"{a}}ssler]{Sun2014}
Zhiqi Sun, Armin Lambacher, and Reinhard F{\"{a}}ssler.
\newblock {Nascent Adhesions: From Fluctuations to a Hierarchical
  Organization}.
\newblock \emph{Current Biology}, 24\penalty0 (17):\penalty0 R801--R803, sep
  2014.
\newblock ISSN 09609822.
\newblock \doi{10.1016/j.cub.2014.07.061}.

\bibitem[Ye et~al.(2013)Ye, Petrich, Anekal, Lefort, Kasirer-Friede, Shattil,
  Ruppert, Moser, F{\"{a}}ssler, and Ginsberg]{Ye2013}
Feng Ye, Brian~G. Petrich, Praju Anekal, Craig~T. Lefort, Ana Kasirer-Friede,
  Sanford~J. Shattil, Raphael Ruppert, Markus Moser, Reinhard F{\"{a}}ssler,
  and Mark~H. Ginsberg.
\newblock {The Mechanism of Kindlin-Mediated Activation of Integrin
  $\alpha$IIb$\beta$3}.
\newblock \emph{Current Biology}, 23\penalty0 (22):\penalty0 2288--2295, nov
  2013.
\newblock ISSN 09609822.
\newblock \doi{10.1016/j.cub.2013.09.050}.

\bibitem[Cluzel et~al.(2005)Cluzel, Saltel, Lussi, Paulhe, Imhof, and
  Wehrle-Haller]{Cluzel2005}
Caroline Cluzel, Frédéric Saltel, Jost Lussi, Frédérique Paulhe, Beat~A
  Imhof, and Bernhard Wehrle-Haller.
\newblock {The mechanisms and dynamics of $\alpha$v$\beta$3 integrin clustering
  in living cells}.
\newblock \emph{The Journal of Cell Biology}, 171\penalty0 (2):\penalty0
  383--392, 2005.
\newblock ISSN 1540-8140.
\newblock \doi{10.1083/jcb.200503017}.

\bibitem[Saltel et~al.(2009)Saltel, Mortier, Hyt{\"{o}}nen, Jacquier,
  Zimmermann, Vogel, Liu, and Wehrle-Haller]{Saltel2009}
Fr{\'{e}}d{\'{e}}ric Saltel, Eva Mortier, Vesa~P. Hyt{\"{o}}nen, Marie-Claude
  Jacquier, Pascale Zimmermann, Viola Vogel, Wei Liu, and Bernhard
  Wehrle-Haller.
\newblock {New PI(4,5)P2- and membrane proximal integrin–binding motifs in
  the talin head control $\beta$3-integrin clustering}.
\newblock \emph{The Journal of Cell Biology}, 187\penalty0 (5):\penalty0
  715--731, nov 2009.
\newblock ISSN 1540-8140.
\newblock \doi{10.1083/jcb.200908134}.

\bibitem[Ellis et~al.(2014)Ellis, Lostchuck, Goult, Bouaouina, Fairchild,
  L{\'{o}}pez-Ceballos, Calderwood, and Tanentzapf]{Ellis2014}
Stephanie~J. Ellis, Emily Lostchuck, Benjamin~T. Goult, Mohamed Bouaouina,
  Michael~J. Fairchild, Pablo L{\'{o}}pez-Ceballos, David~A. Calderwood, and
  Guy Tanentzapf.
\newblock {The Talin Head Domain Reinforces Integrin-Mediated Adhesion by
  Promoting Adhesion Complex Stability and Clustering}.
\newblock \emph{PLoS Genetics}, 10\penalty0 (11):\penalty0 e1004756, nov 2014.
\newblock ISSN 1553-7404.
\newblock \doi{10.1371/journal.pgen.1004756}.

\bibitem[Humphries et~al.(2007)Humphries, Wang, Streuli, Geiger, Humphries, and
  Ballestrem]{Humphries2007}
Jonathan~D. Humphries, Pengbo Wang, Charles Streuli, Benny Geiger, Martin~J.
  Humphries, and Christoph Ballestrem.
\newblock {Vinculin controls focal adhesion formation by direct interactions
  with talin and actin}.
\newblock \emph{The Journal of Cell Biology}, 179\penalty0 (5):\penalty0
  1043--1057, dec 2007.
\newblock ISSN 1540-8140.
\newblock \doi{10.1083/jcb.200703036}.
\newblock URL
  \url{https://rupress.org/jcb/article/179/5/1043/45072/Vinculin-controls-focal-adhesion-formation-by}.

\bibitem[Barczyk et~al.(2010)Barczyk, Carracedo, and Gullberg]{Barczyk2010}
Malgorzata Barczyk, Sergio Carracedo, and Donald Gullberg.
\newblock {Integrins}.
\newblock \emph{Cell and Tissue Research}, 339\penalty0 (1):\penalty0 269--280,
  jan 2010.
\newblock ISSN 0302-766X.
\newblock \doi{10.1007/s00441-009-0834-6}.

\bibitem[Oakes et~al.(2018)Oakes, Bidone, Beckham, Skeeters, {Ramirez-San
  Juan}, Winter, Voth, and Gardel]{Oakes2646}
Patrick~W. Oakes, Tamara~C. Bidone, Yvonne Beckham, Austin~V. Skeeters,
  Guillermina~R. {Ramirez-San Juan}, Stephen~P. Winter, Gregory~A. Voth, and
  Margaret~L. Gardel.
\newblock {Lamellipodium is a myosin-independent mechanosensor}.
\newblock \emph{Proceedings of the National Academy of Sciences}, 115\penalty0
  (11):\penalty0 2646--2651, mar 2018.
\newblock ISSN 0027-8424.
\newblock \doi{10.1073/pnas.1715869115}.

\bibitem[Bachir et~al.(2014)Bachir, Zareno, Moissoglu, Plow, Gratton, and
  Horwitz]{Bachir2014}
Alexia~I. Bachir, Jessica Zareno, Konstadinos Moissoglu, Edward~F. Plow, Enrico
  Gratton, and Alan~R. Horwitz.
\newblock {Integrin-Associated Complexes Form Hierarchically with Variable
  Stoichiometry in Nascent Adhesions}.
\newblock \emph{Current Biology}, 24\penalty0 (16):\penalty0 1845--1853, aug
  2014.
\newblock ISSN 09609822.
\newblock \doi{10.1016/j.cub.2014.07.011}.

\bibitem[Horton et~al.(2015)Horton, Byron, Askari, Ng, Millon-Fr{\'{e}}millon,
  Robertson, Koper, Paul, Warwood, Knight, Humphries, and
  Humphries]{Horton2015}
Edward~R. Horton, Adam Byron, Janet~A. Askari, Daniel H.~J. Ng, Ang{\'{e}}lique
  Millon-Fr{\'{e}}millon, Joseph Robertson, Ewa~J. Koper, Nikki~R. Paul, Stacey
  Warwood, David Knight, Jonathan~D. Humphries, and Martin~J. Humphries.
\newblock {Definition of a consensus integrin adhesome and its dynamics during
  adhesion complex assembly and disassembly}.
\newblock \emph{Nature Cell Biology}, 17\penalty0 (12):\penalty0 1577--1587,
  dec 2015.
\newblock ISSN 1465-7392.
\newblock \doi{10.1038/ncb3257}.

\bibitem[Michael and Parsons(2020)]{Michael2020}
Magdalene Michael and Maddy Parsons.
\newblock {New perspectives on integrin-dependent adhesions}.
\newblock \emph{Current Opinion in Cell Biology}, 63:\penalty0 31--37, apr
  2020.
\newblock ISSN 09550674.
\newblock \doi{10.1016/j.ceb.2019.12.008}.

\bibitem[Humphries et~al.(2019)Humphries, Chastney, Askari, and
  Humphries]{Humphries2019}
Jonathan~D Humphries, Megan~R Chastney, Janet~A Askari, and Martin~J Humphries.
\newblock {Signal transduction via integrin adhesion complexes}.
\newblock \emph{Current Opinion in Cell Biology}, 56:\penalty0 14--21, feb
  2019.
\newblock ISSN 09550674.
\newblock \doi{10.1016/j.ceb.2018.08.004}.

\bibitem[Alday-Parejo et~al.(2019)Alday-Parejo, Stupp, and
  R{\"{u}}egg]{Alday-Parejo2019}
Alday-Parejo, Stupp, and R{\"{u}}egg.
\newblock {Are Integrins Still Practicable Targets for Anti-Cancer Therapy?}
\newblock \emph{Cancers}, 11\penalty0 (7):\penalty0 978, jul 2019.
\newblock ISSN 2072-6694.
\newblock \doi{10.3390/cancers11070978}.

\bibitem[Bachmann et~al.(2019)Bachmann, Kukkurainen, Hyt{\"{o}}nen, and
  Wehrle-Haller]{Bachmann2019}
Michael Bachmann, Sampo Kukkurainen, Vesa~P. Hyt{\"{o}}nen, and Bernhard
  Wehrle-Haller.
\newblock {Cell Adhesion by Integrins}.
\newblock \emph{Physiological Reviews}, 99\penalty0 (4):\penalty0 1655--1699,
  oct 2019.
\newblock ISSN 0031-9333.
\newblock \doi{10.1152/physrev.00036.2018}.

\bibitem[Shimaoka and Springer(2003)]{Shimaoka2003}
Motomu Shimaoka and Timothy~A. Springer.
\newblock {Therapeutic antagonists and conformational regulation of integrin
  function}.
\newblock \emph{Nature Reviews Drug Discovery}, 2\penalty0 (9):\penalty0
  703--716, sep 2003.
\newblock ISSN 1474-1776.
\newblock \doi{10.1038/nrd1174}.

\bibitem[Campbell and Humphries(2011)]{Campbell2011}
I.~D. Campbell and M.~J. Humphries.
\newblock {Integrin Structure, Activation, and Interactions}.
\newblock \emph{Cold Spring Harbor Perspectives in Biology}, 3\penalty0
  (3):\penalty0 a004994--a004994, mar 2011.
\newblock ISSN 1943-0264.
\newblock \doi{10.1101/cshperspect.a004994}.

\bibitem[Hynes(2002)]{Hynes2002}
Richard~O. Hynes.
\newblock {Integrins}.
\newblock \emph{Cell}, 110\penalty0 (6):\penalty0 673--687, sep 2002.
\newblock ISSN 00928674.
\newblock \doi{10.1016/S0092-8674(02)00971-6}.

\bibitem[Xiong(2001)]{Xiong2001}
J.-P. Xiong.
\newblock {Crystal Structure of the Extracellular Segment of Integrin alpha
  Vbeta 3}.
\newblock \emph{Science}, 294\penalty0 (5541):\penalty0 339--345, oct 2001.
\newblock ISSN 00368075.
\newblock \doi{10.1126/science.1064535}.

\bibitem[Moreno-Layseca et~al.(2019)Moreno-Layseca, Icha, Hamidi, and
  Ivaska]{Moreno-Layseca2019}
Paulina Moreno-Layseca, Jaroslav Icha, Hellyeh Hamidi, and Johanna Ivaska.
\newblock {Integrin trafficking in cells and tissues}.
\newblock \emph{Nature Cell Biology}, 21\penalty0 (2):\penalty0 122--132, feb
  2019.
\newblock ISSN 1465-7392.
\newblock \doi{10.1038/s41556-018-0223-z}.

\bibitem[Humphries et~al.(2006)Humphries, Byron, and Humphries]{Humphries2006}
J.~D. Humphries, Adam Byron, and Martin~J. Humphries.
\newblock {Integrin ligands at a glance}.
\newblock \emph{Journal of Cell Science}, 119\penalty0 (19):\penalty0
  3901--3903, oct 2006.
\newblock ISSN 0021-9533.
\newblock \doi{10.1242/jcs.03098}.

\bibitem[Takada et~al.(2007)Takada, Ye, and Simon]{Takada2007}
Yoshikazu Takada, Xiaojing Ye, and Scott Simon.
\newblock {The integrins}.
\newblock \emph{Genome Biology}, 8\penalty0 (5):\penalty0 215, 2007.
\newblock ISSN 14656906.
\newblock \doi{10.1186/gb-2007-8-5-215}.

\bibitem[Byron et~al.(2015)Byron, Askari, Humphries, Jacquemet, Koper, Warwood,
  Choi, Stroud, Chen, Knight, and Humphries]{Byron2015}
Adam Byron, Janet~A. Askari, Jonathan~D. Humphries, Guillaume Jacquemet, Ewa~J.
  Koper, Stacey Warwood, Colin~K. Choi, Matthew~J. Stroud, Christopher~S. Chen,
  David Knight, and Martin~J. Humphries.
\newblock {A proteomic approach reveals integrin activation state-dependent
  control of microtubule cortical targeting}.
\newblock \emph{Nature Communications}, 6\penalty0 (1):\penalty0 6135, may
  2015.
\newblock ISSN 2041-1723.
\newblock \doi{10.1038/ncomms7135}.

\bibitem[Jones et~al.(2015)Jones, Humphries, Byron, Millon‐Fr{\'{e}}millon,
  Robertson, Paul, Ng, Askari, and Humphries]{Jones2015}
Matthew~C. Jones, Jonathan~D. Humphries, Adam Byron, Ang{\'{e}}lique
  Millon‐Fr{\'{e}}millon, Joseph Robertson, Nikki~R. Paul, Daniel H.~J. Ng,
  Janet~A. Askari, and Martin~J. Humphries.
\newblock {Isolation of Integrin‐Based Adhesion Complexes}.
\newblock \emph{Current Protocols in Cell Biology}, 66\penalty0 (1), mar 2015.
\newblock ISSN 1934-2500.
\newblock \doi{10.1002/0471143030.cb0908s66}.

\bibitem[Kuo et~al.(2011)Kuo, Han, Hsiao, {Yates III}, and Waterman]{Kuo2011}
Jean-Cheng Kuo, Xuemei Han, Cheng-Te Hsiao, John~R. {Yates III}, and Clare~M.
  Waterman.
\newblock {Analysis of the myosin-II-responsive focal adhesion proteome reveals
  a role for $\beta$-Pix in negative regulation of focal adhesion maturation}.
\newblock \emph{Nature Cell Biology}, 13\penalty0 (4):\penalty0 383--393, apr
  2011.
\newblock ISSN 1465-7392.
\newblock \doi{10.1038/ncb2216}.

\bibitem[Schiller et~al.(2011)Schiller, Friedel, Boulegue, and
  F{\"{a}}ssler]{Schiller2011}
Herbert~B Schiller, Caroline~C Friedel, Cyril Boulegue, and Reinhard
  F{\"{a}}ssler.
\newblock {Quantitative proteomics of the integrin adhesome show a myosin
  II‐dependent recruitment of LIM domain proteins}.
\newblock \emph{EMBO reports}, 12\penalty0 (3):\penalty0 259--266, mar 2011.
\newblock ISSN 1469-221X.
\newblock \doi{10.1038/embor.2011.5}.

\bibitem[Zaidel-Bar et~al.(2007)Zaidel-Bar, Itzkovitz, Ma'ayan, Iyengar, and
  Geiger]{Zaidel-Bar2007}
Ronen Zaidel-Bar, Shalev Itzkovitz, Avi Ma'ayan, Ravi Iyengar, and Benjamin
  Geiger.
\newblock {Functional atlas of the integrin adhesome}.
\newblock \emph{Nature Cell Biology}, 9\penalty0 (8):\penalty0 858--867, aug
  2007.
\newblock ISSN 1465-7392.
\newblock \doi{10.1038/ncb0807-858}.

\bibitem[Horton et~al.(2016)Horton, Astudillo, Humphries, and
  Humphries]{Horton2016}
Edward~R. Horton, Pablo Astudillo, Martin~J. Humphries, and Jonathan~D.
  Humphries.
\newblock {Mechanosensitivity of integrin adhesion complexes: role of the
  consensus adhesome}.
\newblock \emph{Experimental Cell Research}, 343\penalty0 (1):\penalty0 7--13,
  apr 2016.
\newblock ISSN 00144827.
\newblock \doi{10.1016/j.yexcr.2015.10.025}.

\bibitem[Parad{\v{z}}ik et~al.(2020)Parad{\v{z}}ik, Humphries,
  Stojanovi{\'{c}}, Nesti{\'{c}}, Majhen, Dekani{\'{c}}, Samar{\v{z}}ija,
  Sedda, Weber, Humphries, and Ambriovi{\'{c}}-Ristov]{Paradzik2020}
Mladen Parad{\v{z}}ik, Jonathan~D. Humphries, Nikolina Stojanovi{\'{c}}, Davor
  Nesti{\'{c}}, Dragomira Majhen, Ana Dekani{\'{c}}, Ivana Samar{\v{z}}ija,
  Delphine Sedda, Igor Weber, Martin~J. Humphries, and Andreja
  Ambriovi{\'{c}}-Ristov.
\newblock {KANK2 Links $\alpha$V$\beta$5 Focal Adhesions to Microtubules and
  Regulates Sensitivity to Microtubule Poisons and Cell Migration}.
\newblock \emph{Frontiers in Cell and Developmental Biology}, 8, mar 2020.
\newblock ISSN 2296-634X.
\newblock \doi{10.3389/fcell.2020.00125}.

\bibitem[Sun et~al.(2016)Sun, Tseng, Tan, Senger, Kurzawa, Dedden, Mizuno,
  Wasik, Thery, Dunn, and F{\"{a}}ssler]{Sun2016}
Zhiqi Sun, Hui-Yuan Tseng, Steven Tan, Fabrice Senger, Laetitia Kurzawa, Dirk
  Dedden, Naoko Mizuno, Anita~A. Wasik, Manuel Thery, Alexander~R. Dunn, and
  Reinhard F{\"{a}}ssler.
\newblock {Kank2 activates talin, reduces force transduction across integrins
  and induces central adhesion formation}.
\newblock \emph{Nature Cell Biology}, 18\penalty0 (9):\penalty0 941--953, sep
  2016.
\newblock ISSN 1465-7392.
\newblock \doi{10.1038/ncb3402}.

\bibitem[Stehbens and Wittmann(2012)]{Stehbens2012}
Samantha Stehbens and Torsten Wittmann.
\newblock {Targeting and transport: How microtubules control focal adhesion
  dynamics}.
\newblock \emph{The Journal of Cell Biology}, 198\penalty0 (4):\penalty0
  481--489, aug 2012.
\newblock ISSN 1540-8140.
\newblock \doi{10.1083/jcb.201206050}.

\bibitem[Puklin-Faucher et~al.(2006)Puklin-Faucher, Gao, Schulten, and
  Vogel]{Puklin-Faucher2006}
Eileen Puklin-Faucher, Mu~Gao, Klaus Schulten, and Viola Vogel.
\newblock {How the headpiece hinge angle is opened: new insights into the
  dynamics of integrin activation}.
\newblock \emph{Journal of Cell Biology}, 175\penalty0 (2):\penalty0 349--360,
  oct 2006.
\newblock ISSN 1540-8140.
\newblock \doi{10.1083/jcb.200602071}.

\bibitem[Chen et~al.(2011)Chen, Lou, Hsin, Schulten, Harvey, and Zhu]{Chen2011}
Wei Chen, Jizhong Lou, Jen Hsin, Klaus Schulten, Stephen~C. Harvey, and Cheng
  Zhu.
\newblock {Molecular Dynamics Simulations of Forced Unbending of Integrin
  $\alpha$V$\beta$3}.
\newblock \emph{PLoS Computational Biology}, 7\penalty0 (2):\penalty0 e1001086,
  feb 2011.
\newblock ISSN 1553-7358.
\newblock \doi{10.1371/journal.pcbi.1001086}.

\bibitem[Pagani and Gohlke(2018)]{Pagani2018}
Giulia Pagani and Holger Gohlke.
\newblock {On the contributing role of the transmembrane domain for
  subunit-specific sensitivity of integrin activation}.
\newblock \emph{Scientific Reports}, 8\penalty0 (1):\penalty0 5733, dec 2018.
\newblock ISSN 2045-2322.
\newblock \doi{10.1038/s41598-018-23778-5}.

\bibitem[Kalli et~al.(2017)Kalli, Rog, Vattulainen, Campbell, and
  Sansom]{Kalli2017}
Antreas~C. Kalli, Tomasz Rog, Ilpo Vattulainen, Iain~D. Campbell, and Mark
  S.~P. Sansom.
\newblock {The Integrin Receptor in Biologically Relevant Bilayers: Insights
  from Molecular Dynamics Simulations}.
\newblock \emph{The Journal of Membrane Biology}, 250\penalty0 (4):\penalty0
  337--351, aug 2017.
\newblock ISSN 0022-2631.
\newblock \doi{10.1007/s00232-016-9908-z}.

\bibitem[Shams and Mofrad(2017)]{Shams2017}
Hengameh Shams and Mohammad R~K Mofrad.
\newblock {$\alpha$-Actinin Induces a Kink in the Transmembrane Domain of
  $\beta$3-Integrin and Impairs Activation via Talin}.
\newblock \emph{Biophysical Journal}, 113\penalty0 (4):\penalty0 948--956, aug
  2017.
\newblock ISSN 00063495.
\newblock \doi{10.1016/j.bpj.2017.06.064}.

\bibitem[Desgrosellier and Cheresh(2010)]{Desgrosellier2010}
Jay~S. Desgrosellier and David~A. Cheresh.
\newblock {Integrins in cancer: biological implications and therapeutic
  opportunities}.
\newblock \emph{Nature Reviews Cancer}, 10\penalty0 (1):\penalty0 9--22, jan
  2010.
\newblock ISSN 1474-175X.
\newblock \doi{10.1038/nrc2748}.

\bibitem[Dickreuter and Cordes(2017)]{Dickreuter2017}
Ellen Dickreuter and Nils Cordes.
\newblock {The cancer cell adhesion resistome: mechanisms, targeting and
  translational approaches}.
\newblock \emph{Biological Chemistry}, 398\penalty0 (7):\penalty0 721--735, jun
  2017.
\newblock ISSN 1437-4315.
\newblock \doi{10.1515/hsz-2016-0326}.

\bibitem[Hamidi and Ivaska(2018)]{Hamidi2018}
Hellyeh Hamidi and Johanna Ivaska.
\newblock {Every step of the way: integrins in cancer progression and
  metastasis}.
\newblock \emph{Nature Reviews Cancer}, 18\penalty0 (9):\penalty0 533--548, sep
  2018.
\newblock ISSN 1474-175X.
\newblock \doi{10.1038/s41568-018-0038-z}.

\bibitem[Seguin et~al.(2015)Seguin, Desgrosellier, Weis, and
  Cheresh]{Seguin2015}
Laetitia Seguin, Jay~S. Desgrosellier, Sara~M. Weis, and David~A. Cheresh.
\newblock {Integrins and cancer: regulators of cancer stemness, metastasis, and
  drug resistance}.
\newblock \emph{Trends in Cell Biology}, 25\penalty0 (4):\penalty0 234--240,
  apr 2015.
\newblock ISSN 09628924.
\newblock \doi{10.1016/j.tcb.2014.12.006}.

\bibitem[Tam et~al.(1998)Tam, Sassoli, Jordan, and Nakada]{Tam1998}
Susan~H. Tam, Patricia~M. Sassoli, Robert~E. Jordan, and Marian~T. Nakada.
\newblock {Abciximab (ReoPro, Chimeric 7E3 Fab) Demonstrates Equivalent
  Affinity and Functional Blockade of Glycoprotein IIb/IIIa and $\alpha$ v
  $\beta$ 3 Integrins}.
\newblock \emph{Circulation}, 98\penalty0 (11):\penalty0 1085--1091, sep 1998.
\newblock ISSN 0009-7322.
\newblock \doi{10.1161/01.CIR.98.11.1085}.

\bibitem[Polman et~al.(2006)Polman, O'Connor, Havrdova, Hutchinson, Kappos,
  Miller, Phillips, Lublin, Giovannoni, Wajgt, Toal, Lynn, Panzara, and
  Sandrock]{Polman2006}
Chris~H. Polman, Paul~W. O'Connor, Eva Havrdova, Michael Hutchinson, Ludwig
  Kappos, David~H. Miller, J.~Theodore Phillips, Fred~D. Lublin, Gavin
  Giovannoni, Andrzej Wajgt, Martin Toal, Frances Lynn, Michael~A. Panzara, and
  Alfred~W. Sandrock.
\newblock {A Randomized, Placebo-Controlled Trial of Natalizumab for Relapsing
  Multiple Sclerosis}.
\newblock \emph{New England Journal of Medicine}, 354\penalty0 (9):\penalty0
  899--910, mar 2006.
\newblock ISSN 0028-4793.
\newblock \doi{10.1056/NEJMoa044397}.

\bibitem[Gordon et~al.(2001)Gordon, Lai, Hamilton, Allison, Srivastava,
  Fouweather, Donoghue, Greenlees, Subhani, Amlot, and Pounder]{Gordon2001}
Fiona~H. Gordon, Clement~W.Y. Lai, Mark~I. Hamilton, Miles~C. Allison,
  Emmanuel~D. Srivastava, Marilyn~G. Fouweather, Stephen Donoghue, Carol
  Greenlees, Javaid Subhani, Peter~L. Amlot, and Roy~E. Pounder.
\newblock {A randomized placebo-controlled trial of a humanized monoclonal
  antibody to $\alpha$4 integrin in active crohn's disease}.
\newblock \emph{Gastroenterology}, 121\penalty0 (2):\penalty0 268--274, aug
  2001.
\newblock ISSN 00165085.
\newblock \doi{10.1053/gast.2001.26260}.

\bibitem[Rosario et~al.(2017)Rosario, Dirks, Milch, Parikh, Bargfrede, Wyant,
  Fedyk, and Fox]{Rosario2017}
Maria Rosario, Nathanael~L. Dirks, Catherine Milch, Asit Parikh, Michael
  Bargfrede, Tim Wyant, Eric Fedyk, and Irving Fox.
\newblock {A Review of the Clinical Pharmacokinetics, Pharmacodynamics, and
  Immunogenicity of Vedolizumab}.
\newblock \emph{Clinical Pharmacokinetics}, 56\penalty0 (11):\penalty0
  1287--1301, nov 2017.
\newblock ISSN 0312-5963.
\newblock \doi{10.1007/s40262-017-0546-0}.

\bibitem[Kim et~al.(2018)Kim, Sheppard, and Chapman]{Kim2018}
Kevin~K. Kim, Dean Sheppard, and Harold~A. Chapman.
\newblock {TGF-$\beta$1 Signaling and Tissue Fibrosis}.
\newblock \emph{Cold Spring Harbor Perspectives in Biology}, 10\penalty0
  (4):\penalty0 a022293, apr 2018.
\newblock ISSN 1943-0264.
\newblock \doi{10.1101/cshperspect.a022293}.

\bibitem[McGrath(2015)]{McGrath2015}
John~A. McGrath.
\newblock {Recently Identified Forms of Epidermolysis Bullosa}.
\newblock \emph{Annals of Dermatology}, 27\penalty0 (6):\penalty0 658, 2015.
\newblock ISSN 1013-9087.
\newblock \doi{10.5021/ad.2015.27.6.658}.

\bibitem[Walko et~al.(2015)Walko, Casta{\~{n}}{\'{o}}n, and Wiche]{Walko2015}
Gernot Walko, Maria~J. Casta{\~{n}}{\'{o}}n, and Gerhard Wiche.
\newblock {Molecular architecture and function of the hemidesmosome}.
\newblock \emph{Cell and Tissue Research}, 360\penalty0 (3):\penalty0 529--544,
  jun 2015.
\newblock ISSN 0302-766X.
\newblock \doi{10.1007/s00441-015-2216-6}.

\bibitem[Alon and Etzioni(2003)]{Alon2003}
Ronen Alon and Amos Etzioni.
\newblock {LAD-III, a novel group of leukocyte integrin activation
  deficiencies}.
\newblock \emph{Trends in Immunology}, 24\penalty0 (10):\penalty0 561--566, oct
  2003.
\newblock ISSN 14714906.
\newblock \doi{10.1016/j.it.2003.08.001}.

\bibitem[Hoffmann et~al.(2011)Hoffmann, Ohlsen, and Hauck]{Hoffmann2011}
Christine Hoffmann, Knut Ohlsen, and Christof~R. Hauck.
\newblock {Integrin-mediated uptake of fibronectin-binding bacteria}.
\newblock \emph{European Journal of Cell Biology}, 90\penalty0 (11):\penalty0
  891--896, nov 2011.
\newblock ISSN 01719335.
\newblock \doi{10.1016/j.ejcb.2011.03.001}.

\bibitem[Hussein et~al.(2015)Hussein, Walker, Abdel-Raouf, Desouky, Montasser,
  and Akula]{Hussein2015}
Hosni A.~M. Hussein, Lia~R. Walker, Usama~M. Abdel-Raouf, Sayed~A. Desouky,
  Abdel Khalek~M. Montasser, and Shaw~M. Akula.
\newblock {Beyond RGD: virus interactions with integrins}.
\newblock \emph{Archives of Virology}, 160\penalty0 (11):\penalty0 2669--2681,
  nov 2015.
\newblock ISSN 0304-8608.
\newblock \doi{10.1007/s00705-015-2579-8}.

\bibitem[Calderwood(2004)]{Calderwood2004}
David~A. Calderwood.
\newblock {Integrin activation}.
\newblock \emph{Journal of Cell Science}, 117\penalty0 (5):\penalty0 657--666,
  mar 2004.
\newblock ISSN 0021-9533.
\newblock \doi{10.1242/jcs.01014}.

\bibitem[Luo et~al.(2007)Luo, Carman, and Springer]{Luo2007}
Bing-Hao Luo, Christopher~V. Carman, and Timothy~A. Springer.
\newblock {Structural Basis of Integrin Regulation and Signaling}.
\newblock \emph{Annual Review of Immunology}, 25\penalty0 (1):\penalty0
  619--647, apr 2007.
\newblock ISSN 0732-0582.
\newblock \doi{10.1146/annurev.immunol.25.022106.141618}.

\bibitem[Sun et~al.(2019)Sun, Costell, and F{\"{a}}ssler]{Sun2019}
Zhiqi Sun, Mercedes Costell, and Reinhard F{\"{a}}ssler.
\newblock {Integrin activation by talin, kindlin and mechanical forces}.
\newblock \emph{Nature Cell Biology}, 21\penalty0 (1):\penalty0 25--31, jan
  2019.
\newblock ISSN 1465-7392.
\newblock \doi{10.1038/s41556-018-0234-9}.

\bibitem[Wang et~al.(2018)Wang, Wu, Zhang, Zhong, Sun, Cao, Ge, Li, Zhang, and
  Chen]{Wang2799}
ShiHui Wang, ChenYu Wu, YueBin Zhang, QingLu Zhong, Hao Sun, WenPeng Cao,
  GaoXiang Ge, GuoHui Li, X.~Frank Zhang, and JianFeng Chen.
\newblock {Integrin $\alpha$4$\beta$7 switches its ligand specificity via
  distinct conformer-specific activation}.
\newblock \emph{The Journal of Cell Biology}, 217\penalty0 (8):\penalty0
  2799--2812, aug 2018.
\newblock ISSN 0021-9525.
\newblock \doi{10.1083/jcb.201710022}.

\bibitem[Li et~al.(2017{\natexlab{a}})Li, Su, Xia, Qin, Humphries, Vestweber,
  Caba{\~{n}}as, Lu, and Springer]{Li2017}
Jing Li, Yang Su, Wei Xia, Yan Qin, Martin~J Humphries, Dietmar Vestweber,
  Carlos Caba{\~{n}}as, Chafen Lu, and Timothy~A Springer.
\newblock {Conformational equilibria and intrinsic affinities define integrin
  activation}.
\newblock \emph{The EMBO Journal}, 36\penalty0 (5):\penalty0 629--645, mar
  2017{\natexlab{a}}.
\newblock ISSN 0261-4189.
\newblock \doi{10.15252/embj.201695803}.

\bibitem[Li and Springer(2018)]{Li2018}
Jing Li and Timothy~A. Springer.
\newblock {Energy landscape differences among integrins establish the framework
  for understanding activation}.
\newblock \emph{The Journal of Cell Biology}, 217\penalty0 (1):\penalty0
  397--412, jan 2018.
\newblock ISSN 0021-9525.
\newblock \doi{10.1083/jcb.201701169}.

\bibitem[Bell(1978)]{Bell1978}
G.~I Bell.
\newblock {Models for the specific adhesion of cells to cells}.
\newblock \emph{Science}, 200\penalty0 (4342):\penalty0 618--627, may 1978.
\newblock ISSN 0036-8075.
\newblock \doi{10.1126/science.347575}.

\bibitem[Dembo et~al.(1988)Dembo, Torney, Saxman, and Hammer]{Dembo1988}
M~Dembo, D.~C. Torney, K.~Saxman, and D.~Hammer.
\newblock {The reaction-limited kinetics of membrane-to-surface adhesion and
  detachment}.
\newblock \emph{Proceedings of the Royal Society of London. Series B.
  Biological Sciences}, 234\penalty0 (1274):\penalty0 55--83, jun 1988.
\newblock ISSN 0080-4649.
\newblock \doi{10.1098/rspb.1988.0038}.

\bibitem[Bihr et~al.(2012)Bihr, Seifert, and Smith]{Bihr2012}
Timo Bihr, Udo Seifert, and Ana-Sun{\v{c}}ana Smith.
\newblock {Nucleation of Ligand-Receptor Domains in Membrane Adhesion}.
\newblock \emph{Physical Review Letters}, 109\penalty0 (25):\penalty0 258101,
  dec 2012.
\newblock ISSN 0031-9007.
\newblock \doi{10.1103/PhysRevLett.109.258101}.

\bibitem[Fenz et~al.(2017)Fenz, Bihr, Schmidt, Merkel, Seifert, Sengupta, and
  Smith]{Fenz2017}
Susanne~F. Fenz, Timo Bihr, Daniel Schmidt, Rudolf Merkel, Udo Seifert, Kheya
  Sengupta, and Ana-Sun{\v{c}}ana~Sun{\v{c}}ana Smith.
\newblock {Membrane fluctuations mediate lateral interaction between cadherin
  bonds}.
\newblock \emph{Nature Physics}, 13\penalty0 (9):\penalty0 906--913, sep 2017.
\newblock ISSN 17452481.
\newblock \doi{10.1038/nphys4138}.

\bibitem[Fenz et~al.(2011)Fenz, Smith, Merkel, and Sengupta]{Fenz2011}
Susanne~Franziska Fenz, Ana-Sun{\v{c}}ana Smith, Rudolf Merkel, and Kheya
  Sengupta.
\newblock {Inter-membrane adhesion mediated by mobile linkers: Effect of
  receptor shortage}.
\newblock \emph{Soft Matter}, 7\penalty0 (3):\penalty0 952--962, 2011.
\newblock ISSN 1744-683X.
\newblock \doi{10.1039/C0SM00550A}.

\bibitem[Kim et~al.(2020)Kim, Lee, Jang, Ye, Hong, Petrich, Ulmer, and
  Kim]{Kim2020}
Jiyoon Kim, Joonha Lee, Jiyoung Jang, Feng Ye, Soon~Jun Hong, Brian~G. Petrich,
  Tobias~S. Ulmer, and Chungho Kim.
\newblock {Topological Adaptation of Transmembrane Domains to the
  Force-Modulated Lipid Bilayer Is a Basis of Sensing Mechanical Force}.
\newblock \emph{Current Biology}, 30\penalty0 (9):\penalty0 1614--1625.e5, may
  2020.
\newblock ISSN 09609822.
\newblock \doi{10.1016/j.cub.2020.02.028}.

\bibitem[Bihr et~al.(2015)Bihr, Seifert, and Smith]{Bihr2015}
Timo Bihr, Udo Seifert, and Ana-Sunčana Smith.
\newblock {Multiscale approaches to protein-mediated interactions between
  membranes—relating microscopic and macroscopic dynamics in radially growing
  adhesions}.
\newblock \emph{New Journal of Physics}, 17\penalty0 (8):\penalty0 083016, aug
  2015.
\newblock ISSN 1367-2630.
\newblock \doi{10.1088/1367-2630/17/8/083016}.

\bibitem[Goennenwein et~al.(2003)Goennenwein, Tanaka, Hu, Moroder, and
  Sackmann]{Goennenwein2003}
Stefanie Goennenwein, Motomu Tanaka, Bin Hu, Luis Moroder, and Erich Sackmann.
\newblock {Functional Incorporation of Integrins into Solid Supported Membranes
  on Ultrathin Films of Cellulose: Impact on Adhesion}.
\newblock \emph{Biophysical Journal}, 85\penalty0 (1):\penalty0 646--655, jul
  2003.
\newblock ISSN 00063495.
\newblock \doi{10.1016/S0006-3495(03)74508-1}.

\bibitem[Streicher et~al.(2009)Streicher, Nassoy, B{\"{a}}rmann, Dif,
  Marchi-Artzner, Brochard-Wyart, Spatz, and Bassereau]{streicher2009}
Pia Streicher, Pierre Nassoy, Michael B{\"{a}}rmann, Aur{\'{e}}lien Dif,
  Val{\'{e}}rie Marchi-Artzner, Fran{\c{c}}oise Brochard-Wyart, Joachim Spatz,
  and Patricia Bassereau.
\newblock {Integrin reconstituted in GUVs: A biomimetic system to study initial
  steps of cell spreading}.
\newblock \emph{Biochimica et Biophysica Acta (BBA) - Biomembranes},
  1788\penalty0 (10):\penalty0 2291--2300, oct 2009.
\newblock ISSN 00052736.
\newblock \doi{10.1016/j.bbamem.2009.07.025}.

\bibitem[Smith et~al.(2008)Smith, Sengupta, Goennenwein, Seifert, and
  Sackmann]{Smith2008}
A.-S. Smith, Kheya Sengupta, Stefanie Goennenwein, Udo Seifert, and Erich
  Sackmann.
\newblock {Force-induced growth of adhesion domains is controlled by receptor
  mobility}.
\newblock \emph{Proceedings of the National Academy of Sciences}, 105\penalty0
  (19):\penalty0 6906--6911, may 2008.
\newblock ISSN 0027-8424.
\newblock \doi{10.1073/pnas.0801706105}.

\bibitem[Ye et~al.(2012)Ye, Kim, and Ginsberg]{Ye2012}
Feng Ye, Chungho Kim, and Mark~H. Ginsberg.
\newblock {Reconstruction of integrin activation}.
\newblock \emph{Blood}, 119\penalty0 (1):\penalty0 26--33, jan 2012.
\newblock ISSN 0006-4971.
\newblock \doi{10.1182/blood-2011-04-292128}.

\bibitem[Park and Goda(2016)]{Park2016}
Yun~Kyung Park and Yukiko Goda.
\newblock {Integrins in synapse regulation}.
\newblock \emph{Nature Reviews Neuroscience}, 17\penalty0 (12):\penalty0
  745--756, dec 2016.
\newblock ISSN 1471-003X.
\newblock \doi{10.1038/nrn.2016.138}.

\bibitem[Calderwood et~al.(2013)Calderwood, Campbell, and
  Critchley]{Calderwood2013}
David~A. Calderwood, Iain~D. Campbell, and David~R. Critchley.
\newblock {Talins and kindlins: partners in integrin-mediated adhesion}.
\newblock \emph{Nature Reviews Molecular Cell Biology}, 14\penalty0
  (8):\penalty0 503--517, aug 2013.
\newblock ISSN 1471-0072.
\newblock \doi{10.1038/nrm3624}.

\bibitem[Ye et~al.(2010)Ye, Hu, Taylor, Ratnikov, Bobkov, McLean, Sligar,
  Taylor, and Ginsberg]{Ye2010}
Feng Ye, Guiqing Hu, Dianne Taylor, Boris Ratnikov, Andrey~A. Bobkov, Mark~A.
  McLean, Stephen~G. Sligar, Kenneth~A. Taylor, and Mark~H. Ginsberg.
\newblock {Recreation of the terminal events in physiological integrin
  activation}.
\newblock \emph{The Journal of Cell Biology}, 188\penalty0 (1):\penalty0
  157--173, jan 2010.
\newblock ISSN 1540-8140.
\newblock \doi{10.1083/jcb.200908045}.

\bibitem[Calderwood et~al.(1999)Calderwood, Zent, Grant, Rees, Hynes, and
  Ginsberg]{Calderwood1999}
David~A. Calderwood, Roy Zent, Richard Grant, D.~Jasper~G. Rees, Richard~O.
  Hynes, and Mark~H. Ginsberg.
\newblock {The talin head domain binds to integrin $\beta$ subunit cytoplasmic
  tails and regulates integrin activation}.
\newblock \emph{Journal of Biological Chemistry}, 274\penalty0 (40):\penalty0
  28071--28074, oct 1999.
\newblock ISSN 00219258.
\newblock \doi{10.1074/jbc.274.40.28071}.

\bibitem[Bledzka et~al.(2012)Bledzka, Liu, Xu, Perera, Yadav, Bialkowska, Qin,
  Ma, and Plow]{Bledzka2012}
Kamila Bledzka, Jianmin Liu, Zhen Xu, H.~Dhanuja Perera, Satya~P. Yadav,
  Katarzyna Bialkowska, Jun Qin, Yan-Qing Ma, and Edward~F. Plow.
\newblock {Spatial Coordination of Kindlin-2 with Talin Head Domain in
  Interaction with Integrin $\beta$ Cytoplasmic Tails}.
\newblock \emph{Journal of Biological Chemistry}, 287\penalty0 (29):\penalty0
  24585--24594, jul 2012.
\newblock ISSN 0021-9258.
\newblock \doi{10.1074/jbc.M111.336743}.

\bibitem[Ma et~al.(2008)Ma, Qin, Wu, and Plow]{Ma2008}
Yan-Qing Ma, Jun Qin, Chuanyue Wu, and Edward~F. Plow.
\newblock {Kindlin-2 (Mig-2): a co-activator of $\beta$3 integrins}.
\newblock \emph{Journal of Cell Biology}, 181\penalty0 (3):\penalty0 439--446,
  may 2008.
\newblock ISSN 1540-8140.
\newblock \doi{10.1083/jcb.200710196}.

\bibitem[Harburger et~al.(2009)Harburger, Bouaouina, and
  Calderwood]{Harburger2009}
David~S. Harburger, Mohamed Bouaouina, and David~A. Calderwood.
\newblock {Kindlin-1 and -2 Directly Bind the C-terminal Region of $\beta$
  Integrin Cytoplasmic Tails and Exert Integrin-specific Activation Effects}.
\newblock \emph{Journal of Biological Chemistry}, 284\penalty0 (17):\penalty0
  11485--11497, apr 2009.
\newblock ISSN 0021-9258.
\newblock \doi{10.1074/jbc.M809233200}.

\bibitem[Shi et~al.(2007)Shi, Ma, Tu, Chen, Wu, Fukuda, Qin, Plow, and
  Wu]{Shi2007}
Xiaohua Shi, Yan-Qing Ma, Yizeng Tu, Ka~Chen, Shan Wu, Koichi Fukuda, Jun Qin,
  Edward~F. Plow, and Chuanyue Wu.
\newblock {The MIG-2/Integrin Interaction Strengthens Cell-Matrix Adhesion and
  Modulates Cell Motility}.
\newblock \emph{Journal of Biological Chemistry}, 282\penalty0 (28):\penalty0
  20455--20466, jul 2007.
\newblock ISSN 0021-9258.
\newblock \doi{10.1074/jbc.M611680200}.

\bibitem[Li and Springer(2017)]{Li2017a}
Jing Li and Timothy~A. Springer.
\newblock {Integrin extension enables ultrasensitive regulation by cytoskeletal
  force}.
\newblock \emph{Proceedings of the National Academy of Sciences}, 114\penalty0
  (18):\penalty0 4685--4690, may 2017.
\newblock ISSN 0027-8424.
\newblock \doi{10.1073/pnas.1704171114}.

\bibitem[Strohmeyer et~al.(2017)Strohmeyer, Bharadwaj, Costell, F{\"{a}}ssler,
  and M{\"{u}}ller]{Strohmeyer2017}
Nico Strohmeyer, Mitasha Bharadwaj, Mercedes Costell, Reinhard F{\"{a}}ssler,
  and Daniel~J. M{\"{u}}ller.
\newblock {Fibronectin-bound $\alpha$5$\beta$1 integrins sense load and signal
  to reinforce adhesion in less than a second}.
\newblock \emph{Nature Materials}, 16\penalty0 (12):\penalty0 1262--1270, dec
  2017.
\newblock ISSN 1476-1122.
\newblock \doi{10.1038/nmat5023}.

\bibitem[Kong et~al.(2009)Kong, Garc{\'{i}}a, Mould, Humphries, and
  Zhu]{Kong2009}
Fang Kong, Andr{\'{e}}s~J. Garc{\'{i}}a, A.~Paul Mould, Martin~J. Humphries,
  and Cheng Zhu.
\newblock {Demonstration of catch bonds between an integrin and its ligand}.
\newblock \emph{The Journal of Cell Biology}, 185\penalty0 (7):\penalty0
  1275--1284, jun 2009.
\newblock ISSN 1540-8140.
\newblock \doi{10.1083/jcb.200810002}.

\bibitem[Choi et~al.(2014)Choi, Duke-Cohan, Chen, Liu, Rossy, Tabarin, Ju, Gui,
  Gaus, Zhu, and Reinherz]{Choi2014}
Y.~I. Choi, J.~S. Duke-Cohan, W.~Chen, B.~Liu, J.~Rossy, T.~Tabarin, L.~Ju,
  J.~Gui, K.~Gaus, C.~Zhu, and E.~L. Reinherz.
\newblock {Dynamic control of $\beta$1 integrin adhesion by the plexinD1-sema3E
  axis}.
\newblock \emph{Proceedings of the National Academy of Sciences}, 111\penalty0
  (1):\penalty0 379--384, jan 2014.
\newblock ISSN 0027-8424.
\newblock \doi{10.1073/pnas.1314209111}.

\bibitem[Chen et~al.(2010)Chen, Lou, and Zhu]{Chen2010}
Wei Chen, Jizhong Lou, and Cheng Zhu.
\newblock {Forcing Switch from Short- to Intermediate- and Long-lived States of
  the $\alpha$A Domain Generates LFA-1/ICAM-1 Catch Bonds}.
\newblock \emph{Journal of Biological Chemistry}, 285\penalty0 (46):\penalty0
  35967--35978, nov 2010.
\newblock ISSN 0021-9258.
\newblock \doi{10.1074/jbc.M110.155770}.

\bibitem[Kong et~al.(2013)Kong, Li, Parks, Dumbauld, Garc{\'{i}}a, Mould,
  Humphries, and Zhu]{Kong2013}
Fang Kong, Zhenhai Li, William~M. Parks, David~W. Dumbauld, Andr{\'{e}}s~J.
  Garc{\'{i}}a, A.~Paul Mould, Martin~J. Humphries, and Cheng Zhu.
\newblock {Cyclic Mechanical Reinforcement of Integrin–Ligand Interactions}.
\newblock \emph{Molecular Cell}, 49\penalty0 (6):\penalty0 1060--1068, mar
  2013.
\newblock ISSN 10972765.
\newblock \doi{10.1016/j.molcel.2013.01.015}.

\bibitem[Fiore et~al.(2014)Fiore, Ju, Chen, Zhu, and Barker]{Fiore2014}
Vincent~F. Fiore, Lining Ju, Yunfeng Chen, Cheng Zhu, and Thomas~H. Barker.
\newblock {Dynamic catch of a Thy-1–$\alpha$5$\beta$1+syndecan-4 trimolecular
  complex}.
\newblock \emph{Nature Communications}, 5\penalty0 (1):\penalty0 4886, dec
  2014.
\newblock ISSN 2041-1723.
\newblock \doi{10.1038/ncomms5886}.

\bibitem[Smith and Seifert(2005)]{smith2005b}
Ana-Sun\v{c}ana Smith and Udo Seifert.
\newblock {Force-Induced De-Adhesion of Specifically Bound Vesicles: Strong
  Adhesion in Competition with Tether Extraction}.
\newblock \emph{Langmuir}, 21\penalty0 (24):\penalty0 11357--11367, nov 2005.
\newblock ISSN 0743-7463.
\newblock \doi{10.1021/la051303f}.

\bibitem[Tan et~al.(2020)Tan, Chang, Anderson, Miller, Prahl, Odde, and
  Dunn]{Tan2020}
Steven~J. Tan, Alice~C. Chang, Sarah~M. Anderson, Cayla~M. Miller, Louis~S.
  Prahl, David~J. Odde, and Alexander~R. Dunn.
\newblock {Regulation and dynamics of force transmission at individual
  cell-matrix adhesion bonds}.
\newblock \emph{Science Advances}, 6\penalty0 (20):\penalty0 eaax0317, may
  2020.
\newblock ISSN 2375-2548.
\newblock \doi{10.1126/sciadv.aax0317}.

\bibitem[Sigal et~al.(2018)Sigal, Zhou, and Zhuang]{Sigal2018}
Yaron~M. Sigal, Ruobo Zhou, and Xiaowei Zhuang.
\newblock {Visualizing and discovering cellular structures with
  super-resolution microscopy}.
\newblock \emph{Science}, 361\penalty0 (6405):\penalty0 880--887, aug 2018.
\newblock ISSN 0036-8075.
\newblock \doi{10.1126/science.aau1044}.

\bibitem[Nicovich et~al.(2017)Nicovich, Owen, and Gaus]{Nicovich2017}
Philip~R Nicovich, Dylan~M Owen, and Katharina Gaus.
\newblock {Turning single-molecule localization microscopy into a quantitative
  bioanalytical tool}.
\newblock \emph{Nature Protocols}, 12\penalty0 (3):\penalty0 453--460, mar
  2017.
\newblock ISSN 1754-2189.
\newblock \doi{10.1038/nprot.2016.166}.

\bibitem[Nieves and Owen(2020)]{Nieves2020}
Daniel~J Nieves and Dylan~M. Owen.
\newblock {Analysis methods for interrogating spatial organisation of single
  molecule localisation microscopy data}.
\newblock \emph{The International Journal of Biochemistry {\&} Cell Biology},
  page 105749, apr 2020.
\newblock ISSN 13572725.
\newblock \doi{10.1016/j.biocel.2020.105749}.

\bibitem[Deschout et~al.(2017)Deschout, Lukes, Sharipov, Feletti, Lasser, and
  Radenovic]{Deschout2017}
Hendrik Deschout, Tomas Lukes, Azat Sharipov, Lely Feletti, Theo Lasser, and
  Aleksandra Radenovic.
\newblock {Combining PALM and SOFI for quantitative imaging of focal adhesions
  in living cells}.
\newblock In J{\"{o}}rg Enderlein, Ingo Gregor, Zygmunt~K. Gryczynski, Rainer
  Erdmann, and Felix Koberling, editors, \emph{Single Molecule Spectroscopy and
  Superresolution Imaging X}, volume 10071, page 100710E, feb 2017.
\newblock ISBN 9781510605831.
\newblock \doi{10.1117/12.2252865}.

\bibitem[Li(2003)]{Li2003}
Renhao Li.
\newblock {Activation of Integrin alphaIIbbeta3 by Modulation of Transmembrane
  Helix Associations}.
\newblock \emph{Science}, 300\penalty0 (5620):\penalty0 795--798, may 2003.
\newblock ISSN 00368075.
\newblock \doi{10.1126/science.1079441}.

\bibitem[Schmidt et~al.(2015)Schmidt, Bihr, Fenz, Merkel, Seifert, Sengupta,
  and Smith]{Schmidt2015}
Daniel Schmidt, Timo Bihr, Susanne Fenz, Rudolf Merkel, Udo Seifert, Kheya
  Sengupta, and Ana-Sun{\v{c}}ana Smith.
\newblock {Crowding of receptors induces ring-like adhesions in model
  membranes}.
\newblock \emph{Biochimica et Biophysica Acta (BBA) - Molecular Cell Research},
  1853\penalty0 (11):\penalty0 2984--2991, nov 2015.
\newblock ISSN 01674889.
\newblock \doi{10.1016/j.bbamcr.2015.05.025}.

\bibitem[Wolfenson et~al.(2016)Wolfenson, Meacci, Liu, Stachowiak, Iskratsch,
  Ghassemi, Roca-Cusachs, O'Shaughnessy, Hone, and Sheetz]{Wolfenson2015}
Haguy Wolfenson, Giovanni Meacci, Shuaimin Liu, Matthew~R. Stachowiak, Thomas
  Iskratsch, Saba Ghassemi, Pere Roca-Cusachs, Ben O'Shaughnessy, James Hone,
  and Michael~P. Sheetz.
\newblock {Tropomyosin controls sarcomere-like contractions for rigidity
  sensing and suppressing growth on soft matrices}.
\newblock \emph{Nature Cell Biology}, 18\penalty0 (1):\penalty0 33--42, jan
  2016.
\newblock ISSN 1465-7392.
\newblock \doi{10.1038/ncb3277}.

\bibitem[Spiess et~al.(2018)Spiess, Hernandez-Varas, Oddone, Olofsson, Blom,
  Waithe, Lock, Lakadamyali, and Str{\"{o}}mblad]{Spiess1929}
Matthias Spiess, Pablo Hernandez-Varas, Anna Oddone, Helene Olofsson, Hans
  Blom, Dominic Waithe, John~G. Lock, Melike Lakadamyali, and Staffan
  Str{\"{o}}mblad.
\newblock {Active and inactive $\beta$1 integrins segregate into distinct
  nanoclusters in focal adhesions}.
\newblock \emph{The Journal of Cell Biology}, 217\penalty0 (6):\penalty0
  1929--1940, jun 2018.
\newblock ISSN 0021-9525.
\newblock \doi{10.1083/jcb.201707075}.

\bibitem[Ye et~al.(2014)Ye, Kim, and Kim]{Ye2014}
Feng Ye, Se-Jong Kim, and Chungho Kim.
\newblock {Intermolecular Transmembrane Domain Interactions Activate Integrin
  $\alpha$IIb$\beta$3}.
\newblock \emph{Journal of Biological Chemistry}, 289\penalty0 (26):\penalty0
  18507--18513, jun 2014.
\newblock ISSN 0021-9258.
\newblock \doi{10.1074/jbc.M113.541888}.

\bibitem[Mehrbod and Mofrad(2013)]{Mehrbod2013}
Mehrdad Mehrbod and Mohammad R.~K. Mofrad.
\newblock {Localized Lipid Packing of Transmembrane Domains Impedes Integrin
  Clustering}.
\newblock \emph{PLoS Computational Biology}, 9\penalty0 (3):\penalty0 e1002948,
  mar 2013.
\newblock ISSN 1553-7358.
\newblock \doi{10.1371/journal.pcbi.1002948}.

\bibitem[Changede and Sheetz(2017)]{Changede2017}
Rishita Changede and Michael Sheetz.
\newblock {Integrin and cadherin clusters: A robust way to organize adhesions
  for cell mechanics}.
\newblock \emph{BioEssays}, 39\penalty0 (1):\penalty0 e201600123, jan 2017.
\newblock ISSN 02659247.
\newblock \doi{10.1002/bies.201600123}.

\bibitem[Bidone et~al.(2019)Bidone, Skeeters, Oakes, and Voth]{Bidone2019}
Tamara~C. Bidone, Austin~V. Skeeters, Patrick~W. Oakes, and Gregory~A. Voth.
\newblock {Multiscale model of integrin adhesion assembly}.
\newblock \emph{PLOS Computational Biology}, 15\penalty0 (6):\penalty0
  e1007077, jun 2019.
\newblock ISSN 1553-7358.
\newblock \doi{10.1371/journal.pcbi.1007077}.

\bibitem[Jane{\v{s}} et~al.(2019)Jane{\v{s}}, Stumpf, Schmidt, Seifert, and
  Smith]{Janes2019}
Josip~Augustin Jane{\v{s}}, Henning Stumpf, Daniel Schmidt, Udo Seifert, and
  Ana-Sun{\v{c}}ana Smith.
\newblock {Statistical Mechanics of an Elastically Pinned Membrane: Static
  Profile and Correlations}.
\newblock \emph{Biophysical Journal}, 116\penalty0 (2):\penalty0 283--295, jan
  2019.
\newblock ISSN 00063495.
\newblock \doi{10.1016/j.bpj.2018.12.003}.

\bibitem[Pierres et~al.(2008)Pierres, Benoliel, Touchard, and
  Bongrand]{pierres2008}
Anne Pierres, Anne-Marie Benoliel, Dominique Touchard, and Pierre Bongrand.
\newblock {How Cells Tiptoe on Adhesive Surfaces before Sticking}.
\newblock \emph{Biophysical Journal}, 94\penalty0 (10):\penalty0 4114--4122,
  may 2008.
\newblock ISSN 00063495.
\newblock \doi{10.1529/biophysj.107.125278}.

\bibitem[Perez et~al.(2008)Perez, Tamada, Sheetz, and Nelson]{Perez2008}
Tomas~D. Perez, Masako Tamada, Michael~P. Sheetz, and W.~James Nelson.
\newblock {Immediate-Early Signaling Induced by E-cadherin Engagement and
  Adhesion}.
\newblock \emph{Journal of Biological Chemistry}, 283\penalty0 (8):\penalty0
  5014--5022, feb 2008.
\newblock ISSN 0021-9258.
\newblock \doi{10.1074/jbc.M705209200}.

\bibitem[{Lam Hui} et~al.(2012){Lam Hui}, Wang, Grooman, Wayt, and
  Upadhyaya]{LamHui2012}
King {Lam Hui}, Chenlu Wang, Brian Grooman, Jessica Wayt, and Arpita Upadhyaya.
\newblock {Membrane Dynamics Correlate with Formation of Signaling Clusters
  during Cell Spreading}.
\newblock \emph{Biophysical Journal}, 102\penalty0 (7):\penalty0 1524--1533,
  apr 2012.
\newblock ISSN 00063495.
\newblock \doi{10.1016/j.bpj.2012.02.015}.

\bibitem[Smith and Sackmann(2009)]{smith2009}
Ana-Sunčana Smith and Erich Sackmann.
\newblock {Progress in Mimetic Studies of Cell Adhesion and the
  Mechanosensing}.
\newblock \emph{ChemPhysChem}, 10\penalty0 (1):\penalty0 66--78, jan 2009.
\newblock ISSN 14394235.
\newblock \doi{10.1002/cphc.200800683}.

\bibitem[Brinkerhoff and Linderman(2005)]{Brinkerhoff2005}
Christopher~J. Brinkerhoff and Jennifer~J. Linderman.
\newblock {Integrin Dimerization and Ligand Organization: Key Components in
  Integrin Clustering for Cell Adhesion}.
\newblock \emph{Tissue Engineering}, 11\penalty0 (5-6):\penalty0 865--876, may
  2005.
\newblock ISSN 1076-3279.
\newblock \doi{10.1089/ten.2005.11.865}.

\bibitem[MacKay and Khadra(2019)]{MacKay2019}
Laurent MacKay and Anmar Khadra.
\newblock {Dynamics of Mechanosensitive Nascent Adhesion Formation}.
\newblock \emph{Biophysical Journal}, 117\penalty0 (6):\penalty0 1057--1073,
  sep 2019.
\newblock ISSN 00063495.
\newblock \doi{10.1016/j.bpj.2019.08.004}.

\bibitem[Bruinsma et~al.(2000)Bruinsma, Behrisch, and Sackmann]{bruinsma2000}
Robijn Bruinsma, Almuth Behrisch, and Erich Sackmann.
\newblock {Adhesive switching of membranes: Experiment and theory}.
\newblock \emph{Physical Review E}, 61\penalty0 (4):\penalty0 4253--4267, apr
  2000.
\newblock ISSN 1063-651X.
\newblock \doi{10.1103/PhysRevE.61.4253}.

\bibitem[Paszek et~al.(2014)Paszek, DuFort, Rossier, Bainer, Mouw, Godula,
  Hudak, Lakins, Wijekoon, Cassereau, Rubashkin, Magbanua, Thorn, Davidson,
  Rugo, Park, Hammer, Giannone, Bertozzi, and Weaver]{Paszek2014}
Matthew~J. Paszek, Christopher~C. DuFort, Olivier Rossier, Russell Bainer,
  Janna~K. Mouw, Kamil Godula, Jason~E. Hudak, Jonathon~N. Lakins, Amanda~C.
  Wijekoon, Luke Cassereau, Matthew~G. Rubashkin, Mark~J. Magbanua, Kurt~S.
  Thorn, Michael~W. Davidson, Hope~S. Rugo, John~W. Park, Daniel~A. Hammer,
  Gr{\'{e}}gory Giannone, Carolyn~R. Bertozzi, and Valerie~M. Weaver.
\newblock {The cancer glycocalyx mechanically primes integrin-mediated growth
  and survival}.
\newblock \emph{Nature}, 511\penalty0 (7509):\penalty0 319--325, jun 2014.
\newblock ISSN 0028-0836.
\newblock \doi{10.1038/nature13535}.

\bibitem[Helfrich(1978)]{helfrich1978}
W.~Helfrich.
\newblock {Steric Interaction of Fluid Membranes in Multilayer Systems}.
\newblock \emph{Zeitschrift f{\"{u}}r Naturforschung A}, 33\penalty0
  (3):\penalty0 305, jan 1978.
\newblock ISSN 1865-7109.
\newblock \doi{10.1515/zna-1978-0308}.

\bibitem[Turlier and Betz(2018)]{Turlier2018}
Herv{\'{e}} Turlier and Timo Betz.
\newblock {Fluctuations in Active Membranes}.
\newblock In Patricia Bassereau and Pierre Sens, editors, \emph{Physics of
  Biological Membranes}, pages 581--619. Springer International Publishing,
  Cham, 2018.
\newblock ISBN 978-3-030-00630-3.
\newblock \doi{10.1007/978-3-030-00630-3_21}.

\bibitem[Schvartzman et~al.(2011)Schvartzman, Palma, Sable, Abramson, Hu,
  Sheetz, and Wind]{Schvartzman2011}
Mark Schvartzman, Matteo Palma, Julia Sable, Justin Abramson, Xian Hu,
  Michael~P. Sheetz, and Shalom~J. Wind.
\newblock {Nanolithographic Control of the Spatial Organization of Cellular
  Adhesion Receptors at the Single-Molecule Level}.
\newblock \emph{Nano Letters}, 11\penalty0 (3):\penalty0 1306--1312, mar 2011.
\newblock ISSN 1530-6984.
\newblock \doi{10.1021/nl104378f}.

\bibitem[Jamali et~al.(2013)Jamali, Jamali, and Mofrad]{Jamali2013}
Yousef Jamali, Tahereh Jamali, and Mohammad~R.K. Mofrad.
\newblock {An agent based model of integrin clustering: Exploring the role of
  ligand clustering, integrin homo-oligomerization, integrin–ligand affinity,
  membrane crowdedness and ligand mobility}.
\newblock \emph{Journal of Computational Physics}, 244:\penalty0 264--278, jul
  2013.
\newblock ISSN 00219991.
\newblock \doi{10.1016/j.jcp.2012.09.010}.

\bibitem[Theodosiou et~al.(2016)Theodosiou, Widmaier, B{\"{o}}ttcher, Rognoni,
  Veelders, Bharadwaj, Lambacher, Austen, M{\"{u}}ller, Zent, and
  F{\"{a}}ssler]{Theodosiou2016}
Marina Theodosiou, Moritz Widmaier, Ralph~T. B{\"{o}}ttcher, Emanuel Rognoni,
  Maik Veelders, Mitasha Bharadwaj, Armin Lambacher, Katharina Austen,
  Daniel~J. M{\"{u}}ller, Roy Zent, and Reinhard F{\"{a}}ssler.
\newblock {Kindlin-2 cooperates with talin to activate integrins and induces
  cell spreading by directly binding paxillin}.
\newblock \emph{eLife}, 5\penalty0 (JANUARY2016), jan 2016.
\newblock ISSN 2050-084X.
\newblock \doi{10.7554/eLife.10130}.

\bibitem[Moser et~al.(2008)Moser, Nieswandt, Ussar, Pozgajova, and
  F{\"{a}}ssler]{Moser2008}
Markus Moser, Bernhard Nieswandt, Siegfried Ussar, Miroslava Pozgajova, and
  Reinhard F{\"{a}}ssler.
\newblock {Kindlin-3 is essential for integrin activation and platelet
  aggregation}.
\newblock \emph{Nature Medicine}, 14\penalty0 (3):\penalty0 325--330, mar 2008.
\newblock ISSN 1078-8956.
\newblock \doi{10.1038/nm1722}.

\bibitem[Zhang et~al.(2008)Zhang, Jiang, Cai, Monkley, Critchley, and
  Sheetz]{Zhang2008}
Xian Zhang, Guoying Jiang, Yunfei Cai, Susan~J. Monkley, David~R. Critchley,
  and Michael~P. Sheetz.
\newblock {Talin depletion reveals independence of initial cell spreading from
  integrin activation and traction}.
\newblock \emph{Nature Cell Biology}, 10\penalty0 (9):\penalty0 1062--1068, sep
  2008.
\newblock ISSN 1465-7392.
\newblock \doi{10.1038/ncb1765}.

\bibitem[Li et~al.(2017{\natexlab{b}})Li, Deng, Sun, Yang, Liu, Wang, Zhang,
  Lin, Wu, Wei, and Yu]{Li9349}
Huadong Li, Yi~Deng, Kang Sun, Haibin Yang, Jie Liu, Meiling Wang, Zhang Zhang,
  Jirong Lin, Chuanyue Wu, Zhiyi Wei, and Cong Yu.
\newblock {Structural basis of kindlin-mediated integrin recognition and
  activation}.
\newblock \emph{Proceedings of the National Academy of Sciences}, 114\penalty0
  (35):\penalty0 9349--9354, aug 2017{\natexlab{b}}.
\newblock ISSN 0027-8424.
\newblock \doi{10.1073/pnas.1703064114}.

\bibitem[Kammerer et~al.(2017)Kammerer, Aretz, and F{\"{a}}ssler]{Kammerer9234}
Patricia Kammerer, Jonas Aretz, and Reinhard F{\"{a}}ssler.
\newblock {Lucky kindlin: A cloverleaf at the integrin tail}.
\newblock \emph{Proceedings of the National Academy of Sciences}, 114\penalty0
  (35):\penalty0 9234--9236, aug 2017.
\newblock ISSN 0027-8424.
\newblock \doi{10.1073/pnas.1712471114}.

\bibitem[Golji and Mofrad(2014)]{Golji2014}
Javad Golji and Mohammad~R.K. Mofrad.
\newblock {The Talin Dimer Structure Orientation Is Mechanically Regulated}.
\newblock \emph{Biophysical Journal}, 107\penalty0 (8):\penalty0 1802--1809,
  oct 2014.
\newblock ISSN 00063495.
\newblock \doi{10.1016/j.bpj.2014.08.038}.

\bibitem[Annibale et~al.(2011{\natexlab{a}})Annibale, Vanni, Scarselli,
  Rothlisberger, and Radenovic]{Annibale2011}
Paolo Annibale, Stefano Vanni, Marco Scarselli, Ursula Rothlisberger, and
  Aleksandra Radenovic.
\newblock {Identification of clustering artifacts in photoactivated
  localization microscopy}.
\newblock \emph{Nature Methods}, 8\penalty0 (7):\penalty0 527--528, jul
  2011{\natexlab{a}}.
\newblock ISSN 1548-7091.
\newblock \doi{10.1038/nmeth.1627}.

\bibitem[Annibale et~al.(2011{\natexlab{b}})Annibale, Vanni, Scarselli,
  Rothlisberger, and Radenovic]{Annibale2011a}
Paolo Annibale, Stefano Vanni, Marco Scarselli, Ursula Rothlisberger, and
  Aleksandra Radenovic.
\newblock {Quantitative Photo Activated Localization Microscopy: Unraveling the
  Effects of Photoblinking}.
\newblock \emph{PLoS ONE}, 6\penalty0 (7):\penalty0 e22678, jul
  2011{\natexlab{b}}.
\newblock ISSN 1932-6203.
\newblock \doi{10.1371/journal.pone.0022678}.

\bibitem[Baumgart et~al.(2016)Baumgart, Arnold, Leskovar, Staszek,
  F{\"{o}}lser, Weghuber, Stockinger, and Sch{\"{u}}tz]{Baumgart2016}
Florian Baumgart, Andreas~M Arnold, Konrad Leskovar, Kaj Staszek, Martin
  F{\"{o}}lser, Julian Weghuber, Hannes Stockinger, and Gerhard~J Sch{\"{u}}tz.
\newblock {Varying label density allows artifact-free analysis of
  membrane-protein nanoclusters}.
\newblock \emph{Nature Methods}, 13\penalty0 (8):\penalty0 661--664, aug 2016.
\newblock ISSN 1548-7091.
\newblock \doi{10.1038/nmeth.3897}.

\bibitem[Spahn et~al.(2016)Spahn, Herrmannsd{\"{o}}rfer, Kuner, and
  Heilemann]{Spahn2016}
Christoph Spahn, Frank Herrmannsd{\"{o}}rfer, Thomas Kuner, and Mike Heilemann.
\newblock {Temporal accumulation analysis provides simplified artifact-free
  analysis of membrane-protein nanoclusters}.
\newblock \emph{Nature Methods}, 13\penalty0 (12):\penalty0 963--964, dec 2016.
\newblock ISSN 1548-7091.
\newblock \doi{10.1038/nmeth.4065}.

\bibitem[Pike et~al.(2019)Pike, Khan, Pallini, Thomas, Mund, Ries, Poulter, and
  Styles]{Pike2019}
Jeremy~A Pike, Abdullah~O Khan, Chiara Pallini, Steven~G Thomas, Markus Mund,
  Jonas Ries, Natalie~S Poulter, and Iain~B Styles.
\newblock {Topological data analysis quantifies biological nano-structure from
  single molecule localization microscopy}.
\newblock \emph{Bioinformatics}, oct 2019.
\newblock ISSN 1367-4803.
\newblock \doi{10.1093/bioinformatics/btz788}.

\bibitem[Williamson et~al.(2020)Williamson, Burn, Simoncelli, Griffi{\'{e}},
  Peters, Davis, and Owen]{Williamson2020}
David~J. Williamson, Garth~L. Burn, Sabrina Simoncelli, Juliette Griffi{\'{e}},
  Ruby Peters, Daniel~M. Davis, and Dylan~M. Owen.
\newblock {Machine learning for cluster analysis of localization microscopy
  data}.
\newblock \emph{Nature Communications}, 11\penalty0 (1):\penalty0 1493, dec
  2020.
\newblock ISSN 2041-1723.
\newblock \doi{10.1038/s41467-020-15293-x}.

\bibitem[Dedden et~al.(2019)Dedden, Schumacher, Kelley, Zacharias,
  Biert{\"{u}}mpfel, F{\"{a}}ssler, and Mizuno]{Dedden2019}
Dirk Dedden, Stephanie Schumacher, Charlotte~F. Kelley, Martin Zacharias,
  Christian Biert{\"{u}}mpfel, Reinhard F{\"{a}}ssler, and Naoko Mizuno.
\newblock {The Architecture of Talin1 Reveals an Autoinhibition Mechanism}.
\newblock \emph{Cell}, 179\penalty0 (1):\penalty0 120--131.e13, sep 2019.
\newblock ISSN 00928674.
\newblock \doi{10.1016/j.cell.2019.08.034}.

\bibitem[Atherton et~al.(2020)Atherton, Lausecker, Carisey, Gilmore, Critchley,
  Barsukov, and Ballestrem]{Atherton2020}
Paul Atherton, Franziska Lausecker, Alexandre Carisey, Andrew Gilmore, David
  Critchley, Igor Barsukov, and Christoph Ballestrem.
\newblock {Relief of talin autoinhibition triggers a force-independent
  association with vinculin}.
\newblock \emph{The Journal of cell biology}, 219\penalty0 (1):\penalty0 1--16,
  2020.
\newblock ISSN 15408140.
\newblock \doi{10.1083/jcb.201903134}.

\bibitem[Roca-Cusachs et~al.(2013)Roca-Cusachs, del Rio, Puklin-Faucher,
  Gauthier, Biais, and Sheetz]{Roca-cusachs2013}
Pere Roca-Cusachs, A.~del Rio, Eileen Puklin-Faucher, Nils~C. Gauthier, Nicolas
  Biais, and Michael~P. Sheetz.
\newblock {Integrin-dependent force transmission to the extracellular matrix by
  -actinin triggers adhesion maturation}.
\newblock \emph{Proceedings of the National Academy of Sciences}, 110\penalty0
  (15):\penalty0 E1361--E1370, apr 2013.
\newblock ISSN 0027-8424.
\newblock \doi{10.1073/pnas.1220723110}.

\bibitem[Bharadwaj et~al.(2017)Bharadwaj, Strohmeyer, Colo, Helenius,
  Beerenwinkel, Schiller, F{\"{a}}ssler, and M{\"{u}}ller]{Bharadwaj2017}
Mitasha Bharadwaj, Nico Strohmeyer, Georgina~P Colo, Jonne Helenius, Niko
  Beerenwinkel, Herbert~B Schiller, Reinhard F{\"{a}}ssler, and Daniel~J
  M{\"{u}}ller.
\newblock {$\alpha$V-class integrins exert dual roles on $\alpha$5$\beta$1
  integrins to strengthen adhesion to fibronectin}.
\newblock \emph{Nature Communications}, 8\penalty0 (1):\penalty0 14348, apr
  2017.
\newblock ISSN 2041-1723.
\newblock \doi{10.1038/ncomms14348}.

\bibitem[Schwarz et~al.(2006)Schwarz, Erdmann, and Bischofs]{Schwarz2006}
Ulrich~S. Schwarz, Thorsten Erdmann, and Ilka~B. Bischofs.
\newblock {Focal adhesions as mechanosensors: The two-spring model}.
\newblock \emph{Biosystems}, 83\penalty0 (2-3):\penalty0 225--232, feb 2006.
\newblock ISSN 03032647.
\newblock \doi{10.1016/j.biosystems.2005.05.019}.

\bibitem[Harland et~al.(2011)Harland, Walcott, and Sun]{Harland2011}
Ben Harland, Sam Walcott, and Sean~X. Sun.
\newblock {Adhesion dynamics and durotaxis in migrating cells}.
\newblock \emph{Physical Biology}, 8\penalty0 (1):\penalty0 015011, feb 2011.
\newblock ISSN 1478-3975.
\newblock \doi{10.1088/1478-3975/8/1/015011}.

\bibitem[Walcott and Sun(2010)]{Walcott2010}
Sam Walcott and Sean~X. Sun.
\newblock {A mechanical model of actin stress fiber formation and substrate
  elasticity sensing in adherent cells}.
\newblock \emph{Proceedings of the National Academy of Sciences}, 107\penalty0
  (17):\penalty0 7757--7762, apr 2010.
\newblock ISSN 0027-8424.
\newblock \doi{10.1073/pnas.0912739107}.

\bibitem[Li et~al.(2010)Li, Bhimalapuram, and Dinner]{Li2010}
Ying Li, Prabhakar Bhimalapuram, and Aaron~R. Dinner.
\newblock {Model for how retrograde actin flow regulates adhesion traction
  stresses}.
\newblock \emph{Journal of Physics: Condensed Matter}, 22\penalty0
  (19):\penalty0 194113, may 2010.
\newblock ISSN 0953-8984.
\newblock \doi{10.1088/0953-8984/22/19/194113}.

\bibitem[Zhu and Williams(2000)]{Zhu2000}
Cheng Zhu and Tom~E. Williams.
\newblock {Modeling Concurrent Binding of Multiple Molecular Species in Cell
  Adhesion}.
\newblock \emph{Biophysical Journal}, 79\penalty0 (4):\penalty0 1850--1857, oct
  2000.
\newblock ISSN 00063495.
\newblock \doi{10.1016/S0006-3495(00)76434-4}.

\bibitem[Welf et~al.(2012)Welf, Naik, and Ogunnaike]{Welf2012}
Erik~S. Welf, Ulhas~P. Naik, and Babatunde~A. Ogunnaike.
\newblock {A Spatial Model for Integrin Clustering as a Result of Feedback
  between Integrin Activation and Integrin Binding}.
\newblock \emph{Biophysical Journal}, 103\penalty0 (6):\penalty0 1379--1389,
  sep 2012.
\newblock ISSN 00063495.
\newblock \doi{10.1016/j.bpj.2012.08.021}.

\bibitem[Vicente-Manzanares et~al.(2007)Vicente-Manzanares, Zareno, Whitmore,
  Choi, and Horwitz]{Vicente-Manzanares2007}
Miguel Vicente-Manzanares, Jessica Zareno, Leanna Whitmore, Colin~K. Choi, and
  Alan~F. Horwitz.
\newblock {Regulation of protrusion, adhesion dynamics, and polarity by myosins
  IIA and IIB in migrating cells}.
\newblock \emph{The Journal of Cell Biology}, 176\penalty0 (5):\penalty0
  573--580, feb 2007.
\newblock ISSN 1540-8140.
\newblock \doi{10.1083/jcb.200612043}.

\bibitem[Gardel et~al.(2010)Gardel, Schneider, Aratyn-Schaus, Waterman,
  Aratyn-Schaus, and Waterman]{Gardel2010a}
Margaret~L. Gardel, Ian~C. Schneider, Yvonne Aratyn-Schaus, Clare~M. Waterman,
  Yvonne Aratyn-Schaus, and Clare~M. Waterman.
\newblock {Mechanical Integration of Actin and Adhesion Dynamics in Cell
  Migration}.
\newblock \emph{Annual Review of Cell and Developmental Biology}, 26\penalty0
  (1):\penalty0 315--333, nov 2010.
\newblock ISSN 1081-0706.
\newblock \doi{10.1146/annurev.cellbio.011209.122036}.

\bibitem[Webb et~al.(2004)Webb, Donais, Whitmore, Thomas, Turner, Parsons, and
  Horwitz]{Webb2004}
Donna~J. Webb, Karen Donais, Leanna~A. Whitmore, Sheila~M. Thomas,
  Christopher~E. Turner, J.~Thomas Parsons, and Alan~F. Horwitz.
\newblock {FAK–Src signalling through paxillin, ERK and MLCK regulates
  adhesion disassembly}.
\newblock \emph{Nature Cell Biology}, 6\penalty0 (2):\penalty0 154--161, feb
  2004.
\newblock ISSN 1465-7392.
\newblock \doi{10.1038/ncb1094}.

\bibitem[Lawson and Schlaepfer(2012)]{Lawson2012a}
Christine Lawson and David Schlaepfer.
\newblock {Integrin adhesions}.
\newblock \emph{Cell Adhesion {\&} Migration}, 6\penalty0 (4):\penalty0
  302--306, jul 2012.
\newblock ISSN 1933-6918.
\newblock \doi{10.4161/cam.20488}.

\bibitem[Swaminathan et~al.(2016)Swaminathan, Fischer, and
  Waterman]{Swaminathan2016}
Vinay Swaminathan, R.~S. Fischer, and Clare~M. Waterman.
\newblock {The FAK–Arp2/3 interaction promotes leading edge advance and
  haptosensing by coupling nascent adhesions to lamellipodia actin}.
\newblock \emph{Molecular Biology of the Cell}, 27\penalty0 (7):\penalty0
  1085--1100, apr 2016.
\newblock ISSN 1059-1524.
\newblock \doi{10.1091/mbc.E15-08-0590}.

\bibitem[He et~al.(2017)He, Sakai, Tsukasaki, Watanabe, and Ikebe]{He2017}
Kangmin He, Tsuyoshi Sakai, Yoshikazu Tsukasaki, Tomonobu~M. Watanabe, and
  Mitsuo Ikebe.
\newblock {Myosin X is recruited to nascent focal adhesions at the leading edge
  and induces multi-cycle filopodial elongation}.
\newblock \emph{Scientific Reports}, 7\penalty0 (1):\penalty0 13685, dec 2017.
\newblock ISSN 2045-2322.
\newblock \doi{10.1038/s41598-017-06147-6}.

\bibitem[Small et~al.(2002)Small, Stradal, Vignal, and Rottner]{Small2002}
J.Victor Small, Theresia Stradal, Emmanuel Vignal, and Klemens Rottner.
\newblock {The lamellipodium: where motility begins}.
\newblock \emph{Trends in Cell Biology}, 12\penalty0 (3):\penalty0 112--120,
  mar 2002.
\newblock ISSN 09628924.
\newblock \doi{10.1016/S0962-8924(01)02237-1}.

\bibitem[B{\"{o}}ttcher et~al.(2017)B{\"{o}}ttcher, Veelders, Rombaut, Faix,
  Theodosiou, Stradal, Rottner, Zent, Herzog, and F{\"{a}}ssler]{Bottcher2017}
Ralph~T. B{\"{o}}ttcher, Maik Veelders, Pascaline Rombaut, Jan Faix, Marina
  Theodosiou, Theresa~E. Stradal, Klemens Rottner, Roy Zent, Franz Herzog, and
  Reinhard F{\"{a}}ssler.
\newblock {Kindlin-2 recruits paxillin and Arp2/3 to promote membrane
  protrusions during initial cell spreading}.
\newblock \emph{The Journal of Cell Biology}, 216\penalty0 (11):\penalty0
  3785--3798, nov 2017.
\newblock ISSN 0021-9525.
\newblock \doi{10.1083/jcb.201701176}.

\bibitem[Zhu et~al.(2019)Zhu, Liu, Lu, Yang, Byzova, and Qin]{Zhu2019}
Liang Zhu, Huan Liu, Fan Lu, Jun Yang, Tatiana~V. Byzova, and Jun Qin.
\newblock {Structural Basis of Paxillin Recruitment by Kindlin-2 in Regulating
  Cell Adhesion}.
\newblock \emph{Structure}, 27\penalty0 (11):\penalty0 1686--1697.e5, nov 2019.
\newblock ISSN 09692126.
\newblock \doi{10.1016/j.str.2019.09.006}.

\bibitem[Choi et~al.(2011)Choi, Zareno, Digman, Gratton, and Horwitz]{Choi2011}
Colin~K. Choi, Jessica Zareno, Michelle~A. Digman, Enrico Gratton, and
  Alan~Rick Horwitz.
\newblock {Cross-Correlated Fluctuation Analysis Reveals
  Phosphorylation-Regulated Paxillin-FAK Complexes in Nascent Adhesions}.
\newblock \emph{Biophysical Journal}, 100\penalty0 (3):\penalty0 583--592, feb
  2011.
\newblock ISSN 00063495.
\newblock \doi{10.1016/j.bpj.2010.12.3719}.

\bibitem[Shan et~al.(2009)Shan, Yu, Li, Pan, Zhang, Wang, Chen, and
  Zhu]{Shan2009}
Yongli Shan, Lihou Yu, Yan Li, Youdong Pan, Qiangge Zhang, Fubin Wang, Jianfeng
  Chen, and Xueliang Zhu.
\newblock {Nudel and FAK as Antagonizing Strength Modulators of Nascent
  Adhesions through Paxillin}.
\newblock \emph{PLoS Biology}, 7\penalty0 (5):\penalty0 e1000116, may 2009.
\newblock ISSN 1545-7885.
\newblock \doi{10.1371/journal.pbio.1000116}.

\bibitem[Subauste et~al.(2004)Subauste, Pertz, Adamson, Turner, Junger, and
  Hahn]{Subauste2004}
M.~Cecilia Subauste, Olivier Pertz, Eileen~D. Adamson, Christopher~E. Turner,
  Sachiko Junger, and Klaus~M. Hahn.
\newblock {Vinculin modulation of paxillin–FAK interactions regulates ERK to
  control survival and motility}.
\newblock \emph{The Journal of Cell Biology}, 165\penalty0 (3):\penalty0
  371--381, may 2004.
\newblock ISSN 1540-8140.
\newblock \doi{10.1083/jcb.200308011}.

\bibitem[Lawson et~al.(2012)Lawson, Lim, Uryu, Chen, Calderwood, and
  Schlaepfer]{Lawson2012}
Christine Lawson, Ssang~Taek Lim, Sean Uryu, Xiao~Lei Chen, David~A.
  Calderwood, and David~D. Schlaepfer.
\newblock {FAK promotes recruitment of talin to nascent adhesions to control
  cell motility}.
\newblock \emph{Journal of Cell Biology}, 196\penalty0 (2):\penalty0 223--232,
  jan 2012.
\newblock ISSN 00219525.
\newblock \doi{10.1083/jcb.201108078}.

\bibitem[Smith et~al.(2006)Smith, Lorz, Seifert, and Sackmann]{smith2006a}
A.-S. Smith, B~G Lorz, U~Seifert, and E~Sackmann.
\newblock {Antagonist-Induced Deadhesion of Specifically Adhered Vesicles}.
\newblock \emph{Biophys. J.}, 90:\penalty0 1064--1080, 2006.
\newblock \doi{10.1529/biophysj.105.062166}.

\bibitem[Iwamoto and Calderwood(2015)]{Iwamoto2015}
Daniel~V. Iwamoto and David~A. Calderwood.
\newblock {Regulation of integrin-mediated adhesions}.
\newblock \emph{Current Opinion in Cell Biology}, 36\penalty0 (Figure
  2):\penalty0 41--47, oct 2015.
\newblock ISSN 09550674.
\newblock \doi{10.1016/j.ceb.2015.06.009}.

\bibitem[Betzig et~al.(2006)Betzig, Patterson, Sougrat, Lindwasser, Olenych,
  Bonifacino, Davidson, Lippincott-Schwartz, and Hess]{betzig2006}
Eric Betzig, George~H. Patterson, Rachid Sougrat, O.~Wolf Lindwasser, Scott
  Olenych, Juan~S. Bonifacino, Michael~W. Davidson, Jennifer
  Lippincott-Schwartz, and Harald~F. Hess.
\newblock {Imaging Intracellular Fluorescent Proteins at Nanometer Resolution}.
\newblock \emph{Science}, 313\penalty0 (5793):\penalty0 1642--1645, sep 2006.
\newblock ISSN 0036-8075.
\newblock \doi{10.1126/science.1127344}.

\bibitem[Shroff et~al.(2007)Shroff, Galbraith, Galbraith, White, Gillette,
  Olenych, Davidson, and Betzig]{Shroff2007}
Hari Shroff, C.~G. Galbraith, James~A. Galbraith, Helen White, Jennifer
  Gillette, Scott Olenych, Michael~W. Davidson, and Eric Betzig.
\newblock {Dual-color superresolution imaging of genetically expressed probes
  within individual adhesion complexes}.
\newblock \emph{Proceedings of the National Academy of Sciences}, 104\penalty0
  (51):\penalty0 20308--20313, dec 2007.
\newblock ISSN 0027-8424.
\newblock \doi{10.1073/pnas.0710517105}.

\bibitem[Yu et~al.(2011)Yu, Law, Suryana, Low, and Sheetz]{Yu2011}
C.-h. Yu, Jaslyn Bee~Khuan Law, Mona Suryana, Hong~Yee Low, and Michael~P.
  Sheetz.
\newblock {Early integrin binding to Arg-Gly-Asp peptide activates actin
  polymerization and contractile movement that stimulates outward
  translocation}.
\newblock \emph{Proceedings of the National Academy of Sciences}, 108\penalty0
  (51):\penalty0 20585--20590, dec 2011.
\newblock ISSN 0027-8424.
\newblock \doi{10.1073/pnas.1109485108}.

\bibitem[Nordenfelt et~al.(2017)Nordenfelt, Moore, Mehta, Kalappurakkal,
  Swaminathan, Koga, Lambert, Baker, Waters, Oldenbourg, Tani, Mayor, Waterman,
  and Springer]{Nordenfelt2017}
Pontus Nordenfelt, Travis~I. Moore, Shalin~B. Mehta, Joseph~Mathew
  Kalappurakkal, Vinay Swaminathan, Nobuyasu Koga, Talley~J. Lambert, David
  Baker, Jennifer~C. Waters, Rudolf Oldenbourg, Tomomi Tani, Satyajit Mayor,
  Clare~M. Waterman, and Timothy~A. Springer.
\newblock {Direction of actin flow dictates integrin LFA-1 orientation during
  leukocyte migration}.
\newblock \emph{Nature Communications}, 8\penalty0 (1):\penalty0 2047, dec
  2017.
\newblock ISSN 2041-1723.
\newblock \doi{10.1038/s41467-017-01848-y}.

\bibitem[Huang et~al.(2017)Huang, Bax, Buckley, Weis, and Dunn]{Huang2017}
Derek~L. Huang, Nicolas~A. Bax, Craig~D. Buckley, William~I. Weis, and
  Alexander~R. Dunn.
\newblock {Vinculin forms a directionally asymmetric catch bond with F-actin}.
\newblock \emph{Science}, 357\penalty0 (6352):\penalty0 703--706, aug 2017.
\newblock ISSN 0036-8075.
\newblock \doi{10.1126/science.aan2556}.

\bibitem[Burridge and Guilluy(2016)]{Burridge2016}
Keith Burridge and Christophe Guilluy.
\newblock {Focal adhesions, stress fibers and mechanical tension}.
\newblock \emph{Experimental Cell Research}, 343\penalty0 (1):\penalty0 14--20,
  apr 2016.
\newblock ISSN 00144827.
\newblock \doi{10.1016/j.yexcr.2015.10.029}.

\bibitem[Ciobanasu et~al.(2014)Ciobanasu, Faivre, and {Le
  Clainche}]{Ciobanasu2014}
Corina Ciobanasu, Bruno Faivre, and Christophe {Le Clainche}.
\newblock {Actomyosin-dependent formation of the mechanosensitive
  talin–vinculin complex reinforces actin anchoring}.
\newblock \emph{Nature Communications}, 5\penalty0 (1):\penalty0 3095, may
  2014.
\newblock ISSN 2041-1723.
\newblock \doi{10.1038/ncomms4095}.

\bibitem[Yao et~al.(2015)Yao, Goult, Chen, Cong, Sheetz, and Yan]{Yao2015}
Mingxi Yao, Benjamin~T. Goult, Hu~Chen, Peiwen Cong, Michael~P. Sheetz, and Jie
  Yan.
\newblock {Mechanical activation of vinculin binding to talin locks talin in an
  unfolded conformation}.
\newblock \emph{Scientific Reports}, 4\penalty0 (1):\penalty0 4610, may 2015.
\newblock ISSN 2045-2322.
\newblock \doi{10.1038/srep04610}.

\bibitem[Yao et~al.(2016)Yao, Goult, Klapholz, Hu, Toseland, Guo, Cong, Sheetz,
  and Yan]{Yao2016}
Mingxi Yao, Benjamin~T. Goult, Benjamin Klapholz, Xian Hu, Christopher~P.
  Toseland, Yingjian Guo, Peiwen Cong, Michael~P. Sheetz, and Jie Yan.
\newblock {The mechanical response of talin}.
\newblock \emph{Nature Communications}, 7\penalty0 (1):\penalty0 11966, sep
  2016.
\newblock ISSN 2041-1723.
\newblock \doi{10.1038/ncomms11966}.

\bibitem[Asaro et~al.(2019)Asaro, Lin, and Zhu]{Asaro2019}
Robert~J. Asaro, Kuanpo Lin, and Qiang Zhu.
\newblock {Mechanosensitivity Occurs along the Adhesome's Force Train and
  Affects Traction Stress}.
\newblock \emph{Biophysical Journal}, 117\penalty0 (9):\penalty0 1599--1614,
  nov 2019.
\newblock ISSN 00063495.
\newblock \doi{10.1016/j.bpj.2019.08.039}.

\bibitem[Diaz et~al.(2020)Diaz, Neubauer, Rechenmacher, Kessler, and
  Missirlis]{Diaz2020}
Carolina Diaz, Stefanie Neubauer, Florian Rechenmacher, Horst Kessler, and
  Dimitris Missirlis.
\newblock {Recruitment of $\alpha$ $\nu$ $\beta$ 3 integrin to $\alpha$ 5
  $\beta$ 1 integrin-induced clusters enables focal adhesion maturation and
  cell spreading}.
\newblock \emph{Journal of Cell Science}, 133\penalty0 (1):\penalty0 jcs232702,
  jan 2020.
\newblock ISSN 0021-9533.
\newblock \doi{10.1242/jcs.232702}.

\bibitem[Levet et~al.(2019)Levet, Julien, Galland, Butler, Beghin, Chazeau,
  Hoess, Ries, Giannone, and Sibarita]{Levet2019}
Florian Levet, Guillaume Julien, R{\'{e}}mi Galland, Corey Butler, Anne Beghin,
  Ana{\"{e}}l Chazeau, Philipp Hoess, Jonas Ries, Gr{\'{e}}gory Giannone, and
  Jean-Baptiste Sibarita.
\newblock {A tessellation-based colocalization analysis approach for
  single-molecule localization microscopy}.
\newblock \emph{Nature Communications}, 10\penalty0 (1):\penalty0 2379, dec
  2019.
\newblock ISSN 2041-1723.
\newblock \doi{10.1038/s41467-019-10007-4}.

\bibitem[Xu et~al.(2018)Xu, Braun, R{\"{o}}nnlund, Widengren, Aspenstr{\"{o}}m,
  and Gad]{Xu2018}
Lei Xu, Laura~J. Braun, Daniel R{\"{o}}nnlund, Jerker Widengren, Pontus
  Aspenstr{\"{o}}m, and Annica K.~B. Gad.
\newblock {Nanoscale localization of proteins within focal adhesions indicates
  discrete functional assemblies with selective force‐dependence}.
\newblock \emph{The FEBS Journal}, 285\penalty0 (9):\penalty0 1635--1652, may
  2018.
\newblock ISSN 1742-464X.
\newblock \doi{10.1111/febs.14433}.

\bibitem[Orr{\'{e}} et~al.(2019)Orr{\'{e}}, Rossier, and Giannone]{Orre2019}
Thomas Orr{\'{e}}, Olivier Rossier, and Gr{\'{e}}gory Giannone.
\newblock {The inner life of integrin adhesion sites: From single molecules to
  functional macromolecular complexes}.
\newblock \emph{Experimental Cell Research}, 379\penalty0 (2):\penalty0
  235--244, jun 2019.
\newblock ISSN 00144827.
\newblock \doi{10.1016/j.yexcr.2019.03.036}.

\bibitem[Rust et~al.(2006)Rust, Bates, and Zhuang]{rust2006}
Michael~J. Rust, Mark Bates, and Xiaowei Zhuang.
\newblock {Sub-diffraction-limit imaging by stochastic optical reconstruction
  microscopy (STORM)}.
\newblock \emph{Nature Methods}, 3\penalty0 (10):\penalty0 793--796, oct 2006.
\newblock ISSN 1548-7091.
\newblock \doi{10.1038/nmeth929}.

\bibitem[Dertinger et~al.(2009)Dertinger, Colyer, Iyer, Weiss, and
  Enderlein]{dertinger2009}
T.~Dertinger, R.~Colyer, G.~Iyer, S.~Weiss, and J.~Enderlein.
\newblock {Fast, background-free, 3D super-resolution optical fluctuation
  imaging (SOFI)}.
\newblock \emph{Proceedings of the National Academy of Sciences}, 106\penalty0
  (52):\penalty0 22287--22292, dec 2009.
\newblock ISSN 0027-8424.
\newblock \doi{10.1073/pnas.0907866106}.

\bibitem[Inavalli et~al.(2019)Inavalli, Lenz, Butler, Angibaud, Compans, Levet,
  T{\o}nnesen, Rossier, Giannone, Thoumine, Hosy, Choquet, Sibarita, and
  N{\"{a}}gerl]{Inavalli2019}
V.~V. G.~Krishna Inavalli, Martin~O. Lenz, Corey Butler, Julie Angibaud,
  Benjamin Compans, Florian Levet, Jan T{\o}nnesen, Olivier Rossier, Gregory
  Giannone, Olivier Thoumine, Eric Hosy, Daniel Choquet, Jean-Baptiste
  Sibarita, and U.~Valentin N{\"{a}}gerl.
\newblock {A super-resolution platform for correlative live single-molecule
  imaging and STED microscopy}.
\newblock \emph{Nature Methods}, 16\penalty0 (12):\penalty0 1263--1268, dec
  2019.
\newblock ISSN 1548-7091.
\newblock \doi{10.1038/s41592-019-0611-8}.

\bibitem[Rossier et~al.(2012)Rossier, Octeau, Sibarita, Leduc, Tessier, Nair,
  Gatterdam, Destaing, Albig{\`{e}}s-Rizo, Tamp{\'{e}}, Cognet, Choquet,
  Lounis, and Giannone]{Rossier2012}
Olivier Rossier, Vivien Octeau, Jean-baptiste Sibarita, C{\'{e}}cile Leduc,
  B{\'{e}}atrice Tessier, Deepak Nair, Volker Gatterdam, Olivier Destaing,
  Corinne Albig{\`{e}}s-Rizo, Robert Tamp{\'{e}}, Laurent Cognet, Daniel
  Choquet, Brahim Lounis, and Gr{\'{e}}gory Giannone.
\newblock {Integrins $\beta$1 and $\beta$3 exhibit distinct dynamic nanoscale
  organizations inside focal adhesions}.
\newblock \emph{Nature Cell Biology}, 14\penalty0 (10):\penalty0 1057--1067,
  oct 2012.
\newblock ISSN 1465-7392.
\newblock \doi{10.1038/ncb2588}.

\bibitem[Balzarotti et~al.(2017)Balzarotti, Eilers, Gwosch, Gynn{\aa},
  Westphal, Stefani, Elf, and Hell]{Balzarotti2017}
Francisco Balzarotti, Yvan Eilers, Klaus~C. Gwosch, Arvid~H. Gynn{\aa}, Volker
  Westphal, Fernando~D. Stefani, Johan Elf, and Stefan~W. Hell.
\newblock {Nanometer resolution imaging and tracking of fluorescent molecules
  with minimal photon fluxes}.
\newblock \emph{Science}, 355\penalty0 (6325):\penalty0 606--612, feb 2017.
\newblock ISSN 0036-8075.
\newblock \doi{10.1126/science.aak9913}.

\bibitem[Sengupta and Smith(2018)]{Sengupta2018}
Kheya Sengupta and Ana-Sun{\v{c}}ana Smith.
\newblock \emph{Adhesion of Biological Membranes}, pages 499--535.
\newblock Springer International Publishing, Cham, 2018.
\newblock ISBN 978-3-030-00630-3.
\newblock \doi{10.1007/978-3-030-00630-3_18}.

\bibitem[Liu et~al.(2014)Liu, Medda, Liu, Galior, Yehl, Spatz, Cavalcanti-Adam,
  and Salaita]{Liu2014}
Yang Liu, Rebecca Medda, Zheng Liu, Kornelia Galior, Kevin Yehl, Joachim~P.
  Spatz, Elisabetta~Ada Cavalcanti-Adam, and Khalid Salaita.
\newblock {Nanoparticle Tension Probes Patterned at the Nanoscale: Impact of
  Integrin Clustering on Force Transmission}.
\newblock \emph{Nano Letters}, 14\penalty0 (10):\penalty0 5539--5546, oct 2014.
\newblock ISSN 1530-6984.
\newblock \doi{10.1021/nl501912g}.

\bibitem[Changede et~al.(2019)Changede, Cai, Wind, and Sheetz]{Changede2019}
Rishita Changede, Haogang Cai, Shalom~J. Wind, and Michael~P. Sheetz.
\newblock {Integrin nanoclusters can bridge thin matrix fibres to form
  cell–matrix adhesions}.
\newblock \emph{Nature Materials}, 18\penalty0 (12):\penalty0 1366--1375, dec
  2019.
\newblock ISSN 1476-1122.
\newblock \doi{10.1038/s41563-019-0460-y}.

\bibitem[Weiss et~al.(2018)Weiss, Frohnmayer, Benk, Haller, Janiesch, Heitkamp,
  B{\"{o}}rsch, Lira, Dimova, Lipowsky, Bodenschatz, Baret, Vidakovic-Koch,
  Sundmacher, Platzman, and Spatz]{Weiss2017}
Marian Weiss, Johannes~Patrick Frohnmayer, Lucia~Theresa Benk, Barbara Haller,
  Jan-Willi Janiesch, Thomas Heitkamp, Michael B{\"{o}}rsch, Rafael~B. Lira,
  Rumiana Dimova, Reinhard Lipowsky, Eberhard Bodenschatz, Jean-Christophe
  Baret, Tanja Vidakovic-Koch, Kai Sundmacher, Ilia Platzman, and Joachim~P.
  Spatz.
\newblock {Sequential bottom-up assembly of mechanically stabilized synthetic
  cells by microfluidics}.
\newblock \emph{Nature Materials}, 17\penalty0 (1):\penalty0 89--96, jan 2018.
\newblock ISSN 1476-1122.
\newblock \doi{10.1038/nmat5005}.

\bibitem[Lock et~al.(2018)Lock, Jones, Askari, Gong, Oddone, Olofsson,
  G{\"{o}}ransson, Lakadamyali, Humphries, and Str{\"{o}}mblad]{Lock2018}
John~G. Lock, Matthew~C. Jones, Janet~A. Askari, Xiaowei Gong, Anna Oddone,
  Helene Olofsson, Sara G{\"{o}}ransson, Melike Lakadamyali, Martin~J.
  Humphries, and Staffan Str{\"{o}}mblad.
\newblock {Reticular adhesions are a distinct class of cell-matrix adhesions
  that mediate attachment during mitosis}.
\newblock \emph{Nature Cell Biology}, 20\penalty0 (11):\penalty0 1290--1302,
  nov 2018.
\newblock ISSN 1465-7392.
\newblock \doi{10.1038/s41556-018-0220-2}.

\bibitem[Elkhatib et~al.(2017)Elkhatib, Bresteau, Baschieri, Rioja, van Niel,
  Vassilopoulos, and Montagnac]{Elkhatib2017}
Nadia Elkhatib, Enzo Bresteau, Francesco Baschieri, Alba~L{\'{o}}pez Rioja,
  Guillaume van Niel, St{\'{e}}phane Vassilopoulos, and Guillaume Montagnac.
\newblock {Tubular clathrin/AP-2 lattices pinch collagen fibers to support 3D
  cell migration}.
\newblock \emph{Science}, 356\penalty0 (6343):\penalty0 eaal4713, jun 2017.
\newblock ISSN 0036-8075.
\newblock \doi{10.1126/science.aal4713}.

\bibitem[Grove et~al.(2014)Grove, Metcalf, Knight, Wavre-Shapton, Sun,
  Protonotarios, Griffin, Lippincott-Schwartz, and Marsh]{Grove2014}
Joe Grove, Daniel~J. Metcalf, Alex~E. Knight, Sil{\`{e}}ne~T. Wavre-Shapton,
  Tony Sun, Emmanouil~D. Protonotarios, Lewis~D. Griffin, Jennifer
  Lippincott-Schwartz, and Mark Marsh.
\newblock {Flat clathrin lattices: stable features of the plasma membrane}.
\newblock \emph{Molecular Biology of the Cell}, 25\penalty0 (22):\penalty0
  3581--3594, nov 2014.
\newblock ISSN 1059-1524.
\newblock \doi{10.1091/mbc.e14-06-1154}.

\bibitem[Leyton-Puig et~al.(2017)Leyton-Puig, Isogai, Argenzio, van~den Broek,
  Klarenbeek, Janssen, Jalink, and Innocenti]{Leyton-Puig2017}
Daniela Leyton-Puig, Tadamoto Isogai, Elisabetta Argenzio, Bram van~den Broek,
  Jeffrey Klarenbeek, Hans Janssen, Kees Jalink, and Metello Innocenti.
\newblock {Flat clathrin lattices are dynamic actin-controlled hubs for
  clathrin-mediated endocytosis and signalling of specific receptors}.
\newblock \emph{Nature Communications}, 8\penalty0 (1):\penalty0 16068, dec
  2017.
\newblock ISSN 2041-1723.
\newblock \doi{10.1038/ncomms16068}.

\bibitem[Lock et~al.(2019)Lock, Baschieri, Jones, Humphries, Montagnac,
  Str{\"{o}}mblad, and Humphries]{Lock2019}
John~G. Lock, Francesco Baschieri, Matthew~C. Jones, Jonathan~D. Humphries,
  Guillaume Montagnac, Staffan Str{\"{o}}mblad, and Martin~J. Humphries.
\newblock {Clathrin-containing adhesion complexes}.
\newblock \emph{Journal of Cell Biology}, 218\penalty0 (7):\penalty0
  2086--2095, jul 2019.
\newblock ISSN 0021-9525.
\newblock \doi{10.1083/jcb.201811160}.

\end{thebibliography}

\end{document}